\documentclass{aa}
\usepackage{graphicx}

\def\arcsec{\hbox{$^{\prime\prime}$}}

\begin{document}


\title{New distances of unresolved dwarf elliptical galaxies in the vicinity 
of the Local Group
\thanks{Based on observations collected at the Nordic Optical Telescope}
}

\author{Rami Rekola\,\inst{1}
\and Helmut Jerjen\,\inst{2}
\and Chris Flynn\,\inst{1}
} 

\offprints{R.~Rekola, email:~rareko@utu.fi}   

\institute{
Tuorla Observatory, VISPA, University of Turku, V\"ais\"al\"antie 20, 
FI-21500 Piikki\"o, Finland
\and 
Research School of Astronomy and Astrophysics, The Australian National 
University, Mt Stromlo Observatory, Cotter Road, Weston ACT 2611, Australia
}

\date{Received ; accepted }

\abstract{We present Surface Brightness Fluctuation distances of nine 
early-type dwarf galaxies and the S0 galaxy NGC 4150 in the Local Volume 
based on deep $B$- and $R$-band CCD images obtained with the 2.56 metre 
Nordic Optical Telescope. Typically, six stellar fields at various 
galactocentric distances have been chosen for each galaxy as appropriately 
free of foreground stars and other contaminants, and Fourier analysed to 
determine the distances, which are found to lie in the range of 3 to 
16\,Mpc. The SBF method is thus demonstrated to efficiently measure 
distances from the ground with mid-aperture telescopes for galaxies for 
which only the tip of the red giant branch method in combination with the 
Hubble Space Telescope has been available until now. We obtained the 
following distance moduli: 
$28.11 \pm 0.15$\,mag (or $4.2 \pm 0.3$\,Mpc) for UGC 1703, 
$27.61 \pm 0.17$\,mag (or $3.3 \pm 0.3$\,Mpc) for KDG 61, 
$29.00 \pm 0.27$\,mag (or $6.3 \pm 0.8$\,Mpc) for UGCA 200, 
$27.74 \pm 0.18$\,mag (or $3.5 \pm 0.3$\,Mpc) for UGC 5442, 
$30.22 \pm 0.17$\,mag (or $11.1 \pm 0.9$\,Mpc) for UGC 5944, 
$30.79 \pm 0.11$\,mag (or $14.4 \pm 0.7$\,Mpc) for NGC 4150, 
$31.02 \pm 0.25$\,mag (or $16.0 \pm 1.9$\,Mpc) for BTS 128, 
$29.27 \pm 0.16$\,mag (or $7.1 \pm 0.6$\,Mpc) for UGC 7639, 
$30.19 \pm 0.23$\,mag (or $10.9 \pm 1.2$\,Mpc) for UGC 8799 
with an alternative distance of $30.61 \pm 0.26$\,mag (or $13.2 \pm 1.7$\,Mpc), 
and $29.60 \pm 0.20$\,mag (or $8.3 \pm 0.8$\,Mpc) for UGC 8882. 
\keywords{ 
galaxies: clusters: individual: M81 group, M101 group, 
CVn I cloud, CVn II cloud, Leo I group, Coma I group -- 
galaxies: dwarf -- 
galaxies: stellar content -- 
galaxies: structure --
galaxies: distances and redshifts} }

\titlerunning{New SBF Distances for Nearby Early-type Galaxies}
\authorrunning{Rekola et al.}

\maketitle

\section{Introduction}
Recent imaging surveys of the Local Group neighbourhood have found and 
identified a large number of low surface brightness galaxy candidates, which 
potentially could be nearby dwarf galaxy systems (e.g.~C\^ot\'e et al.~1997; 
Karachentseva \& Karachentsev 1998; Jerjen et al.~2000a). Many are dwarf 
elliptical galaxies (dEs) located in galaxy groups as satellites of giant 
galaxies. Whether dEs are local or in the background is an important question 
to be answered. For example Moore et al.~(1999) have numerically studied 
galactic and cluster halo substructures in a hierarchical universe, and found 
that simulations predict that the Milky Way galaxy would have many more dwarf 
satellites than are actually observed. This conflict is a strong motivation 
for catalogueing and obtaining accurate distances to dwarf galaxy population 
in the local universe. As a new and numerous target group in extragalactic 
studies dEs serve also to test the peculiar linearity and smoothness of the 
Hubble flow in the local volume. As such they may help to define, or at least 
refine, the dark energy solution to the problem (Baryshev et al.~2001). 

Determining distances to galaxies of this type has been a challenge. They have 
only very little or no neutral hydrogen gas, preventing their detection in 
the radio at 21\,cm and their low surface brightness makes optical 
spectroscopy feasible only for the few brightest objects (Jerjen et 
al.~2000b). Instead, distances must be estimated from their stellar content. 
Taking advantage of the absence of the atmosphere, the Hubble Space Telescope 
(HST) is used to resolve dEs into stars and to measure the tip of the red 
giant branch (TRGB) magnitude (Karachentsev et al.~2000). However, this method 
is expensive in terms of integration time and becomes progressively difficult 
beyond a few Mpc due to crowding effects. 

The Surface Brightness Fluctuation method can be applied to unresolved 
galaxies and thus offers an alternative to efficiently measure distances of 
dEs out to 10\,Mpc and beyond (e.g.~Jerjen 2003). The method was originally 
introduced by Tonry \& Schneider (1988) to measure distances to high surface 
brightness elliptical galaxies. For low surface brightness dEs the method was 
developed (Jerjen et al.~1998; Jerjen et al.~2000b) and calibrated (Jerjen et 
al.~2001) only recently. It is based on the discrete sampling of the 
{\it unresolved} stellar population of a galaxy with the CCD detector and the 
resulting pixel-to-pixel variance due to the statistical noise in numbers of 
red giant branch (RGB) stars.

We can report here on new SBF distances for 10 nearby galaxies as part of our 
continuing project to map the galaxy groups and clouds beyond the Local Group 
out to $\approx$10\,Mpc. We have studied nine dEs and one S0 type of galaxy in 
the northern hemisphere. In Table~\ref{tbl1} we give a complete list of our 
galaxy sample including galaxy name, associated environment, morphological 
type within the extended Hubble classification system (Sandage \& Binggeli 
1984), and coordinates. UGC 1703 is a field galaxy in the direction of NGC 
784, KDG 61 and UGC 5442 are members of the M81 group, UGCA 200 has been 
assumed to be a companion of NGC 3115 but we cannot confirm this assumption, 
UGC 5944 is a member of the Leo I group, NGC 4150 (the S0 galaxy) lies in the 
direction of the Canes Venatici I cloud (CVn I) -- probably behind it, BTS 
128 is a member of the Coma I group, UGC 7639 is a member of the Canes 
Venatici II cloud (CVn II), UGC 8882 is a member of the M101 group, and 
UGC 8799 may lie at the outskirts of the Virgo I cluster. 

\begin{figure*}
\centering
\includegraphics[width=4.4cm]{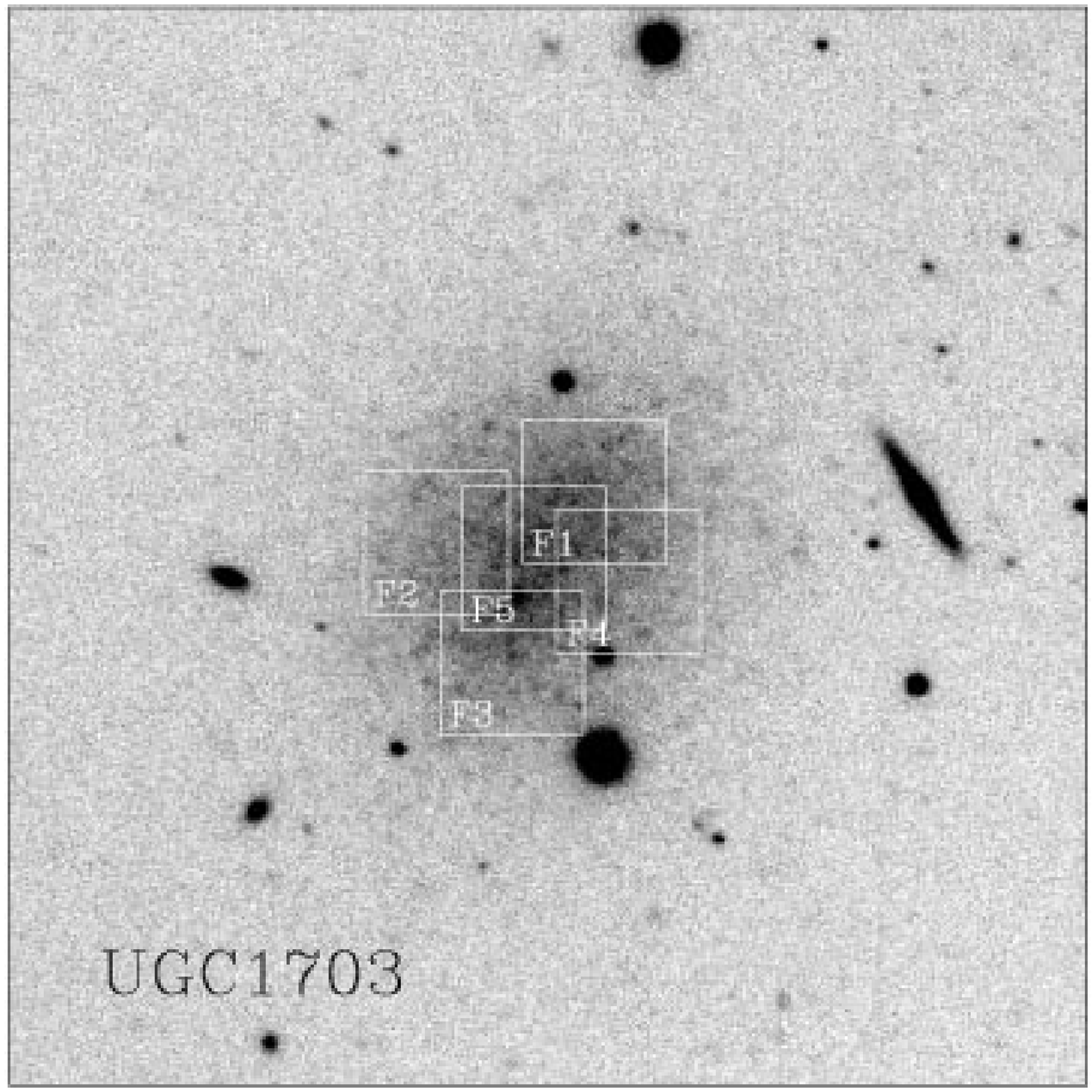}
\includegraphics[width=4.4cm]{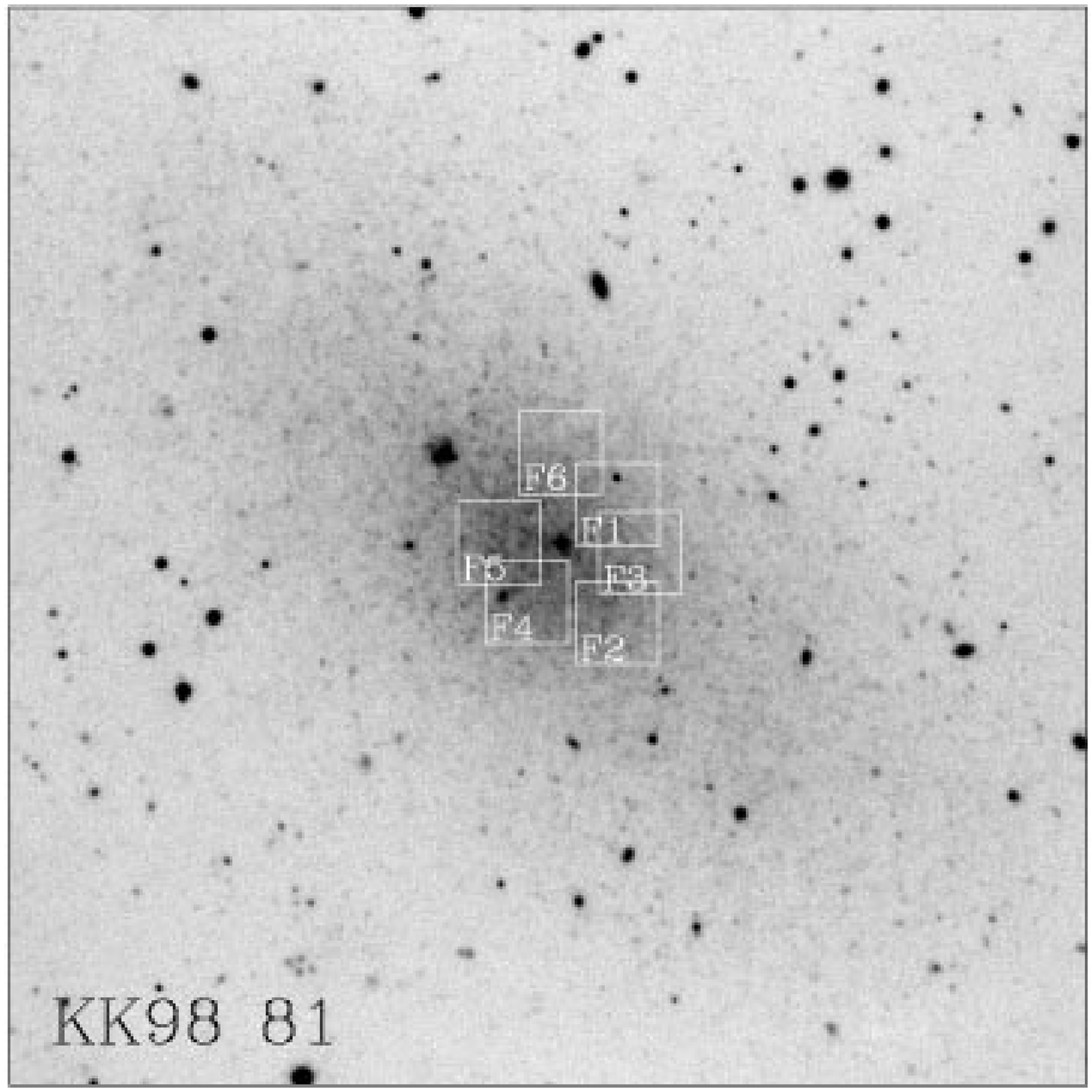}
\includegraphics[width=4.4cm]{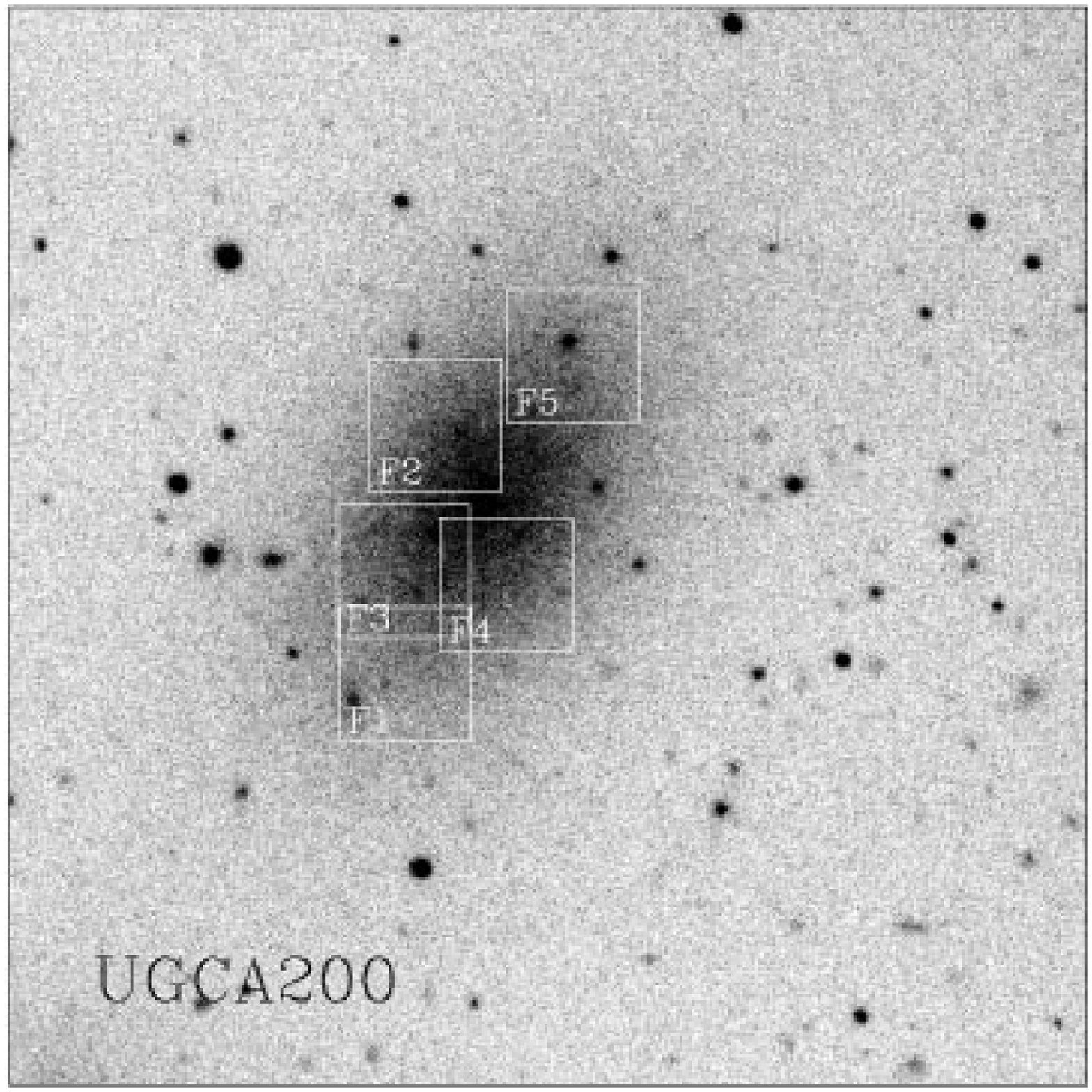}
\includegraphics[width=4.4cm]{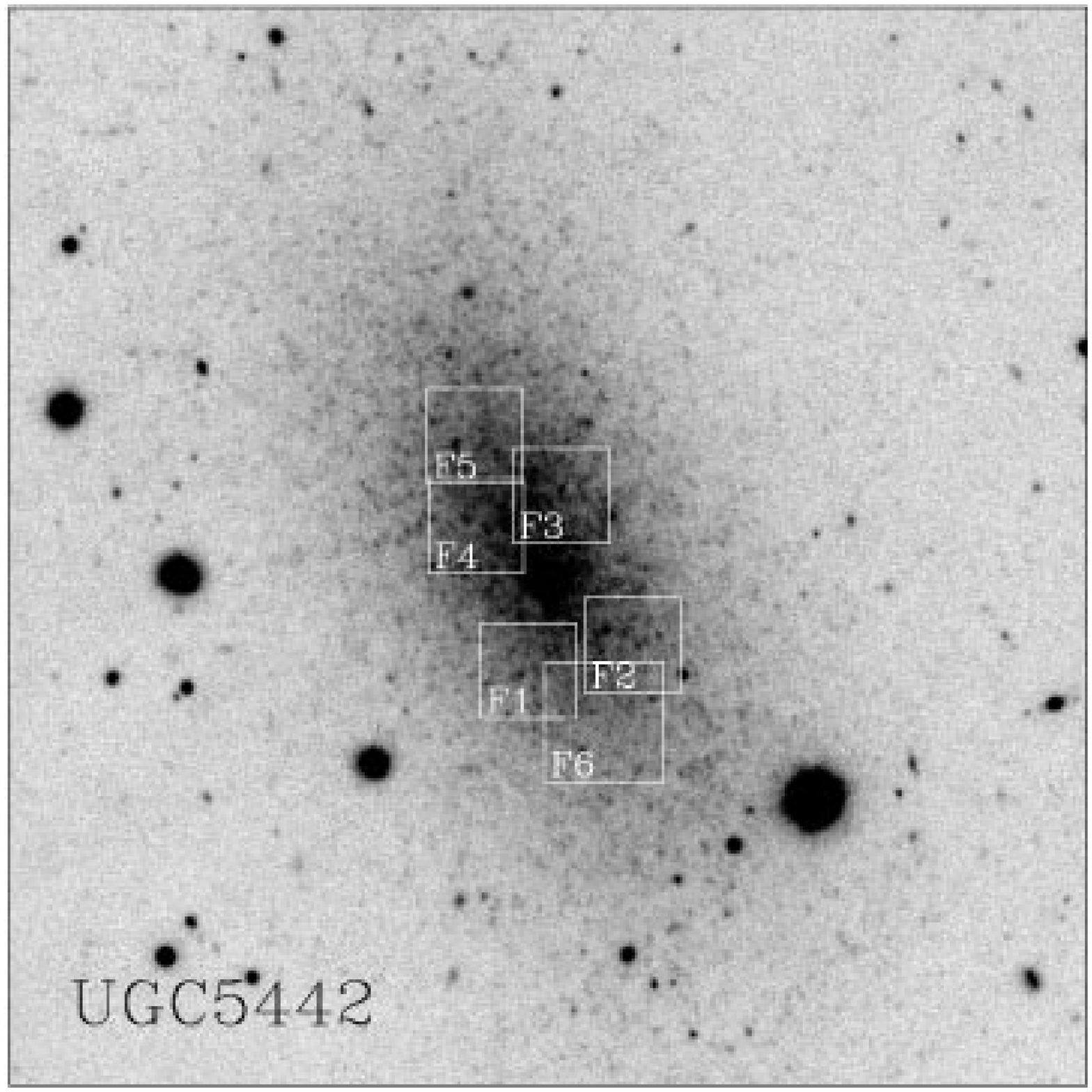}\\
\includegraphics[width=4.4cm]{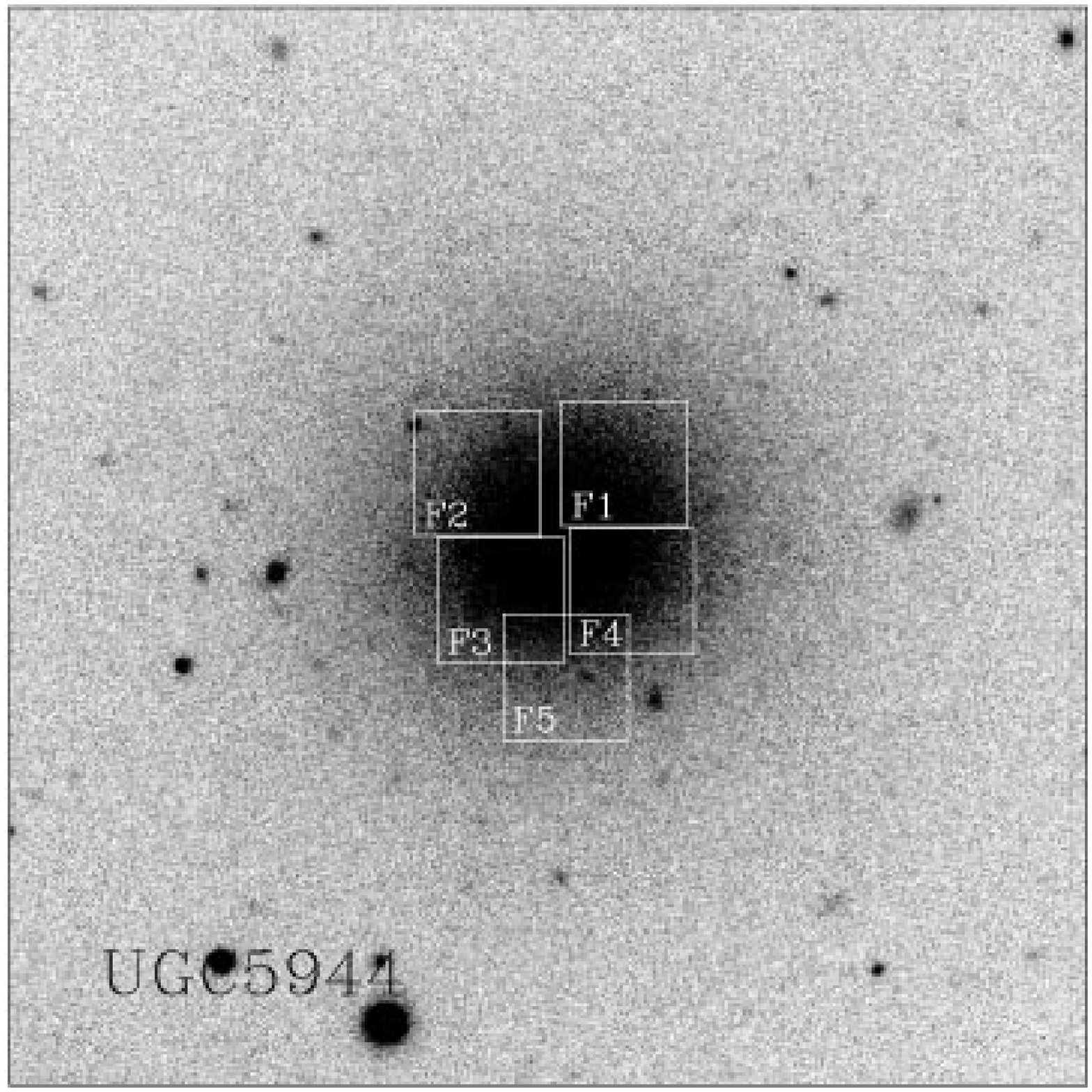}
\includegraphics[width=4.4cm]{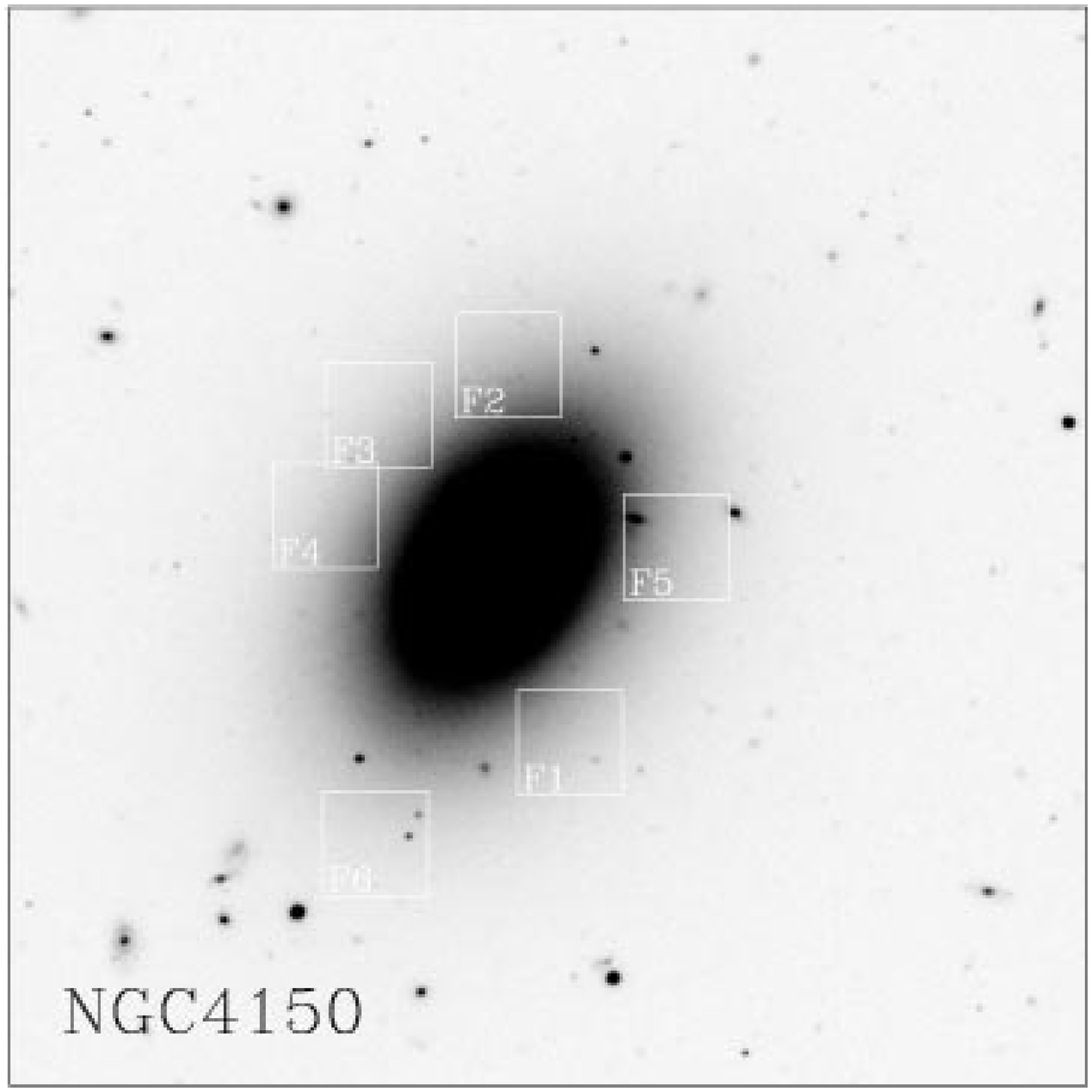}
\includegraphics[width=4.4cm]{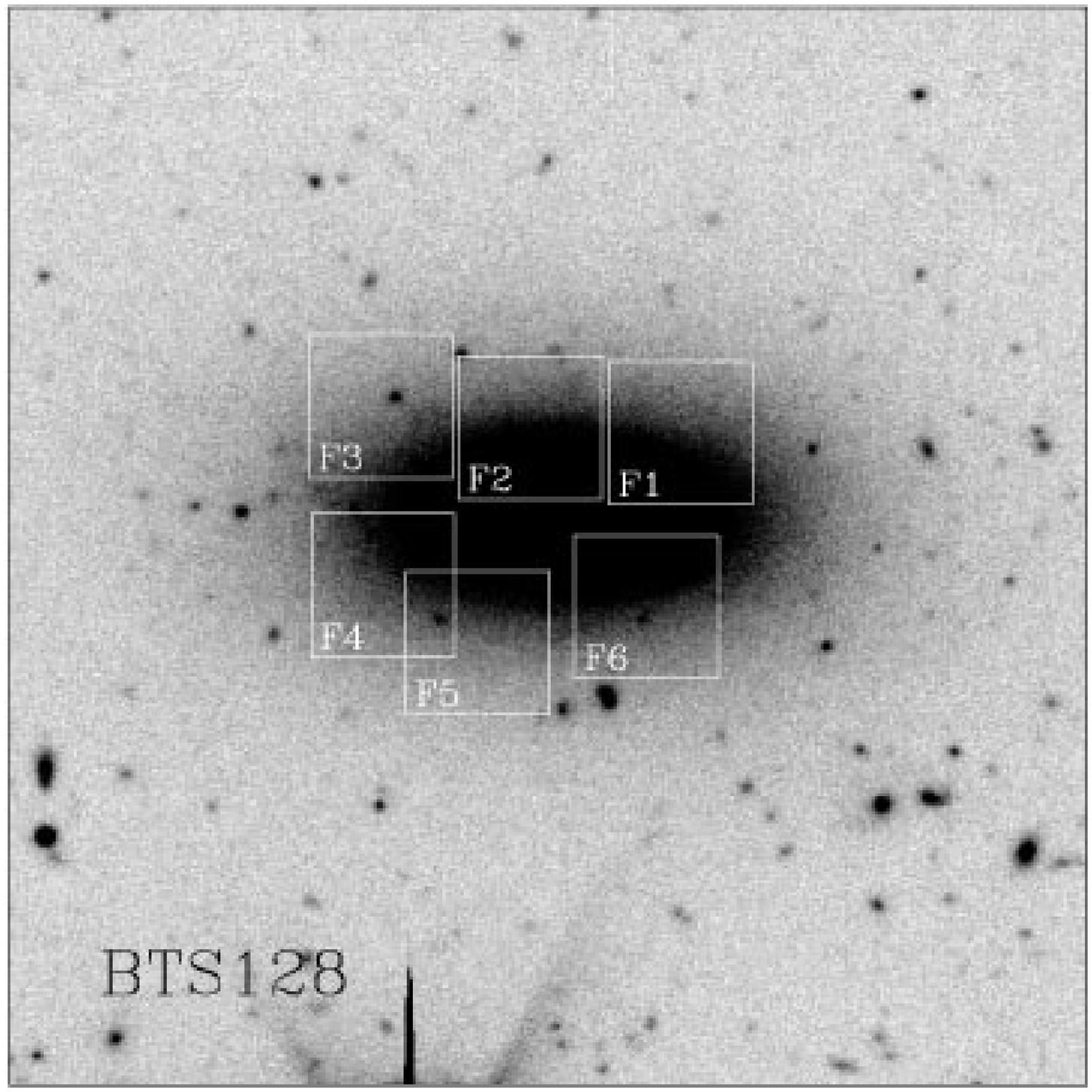}
\includegraphics[width=4.4cm]{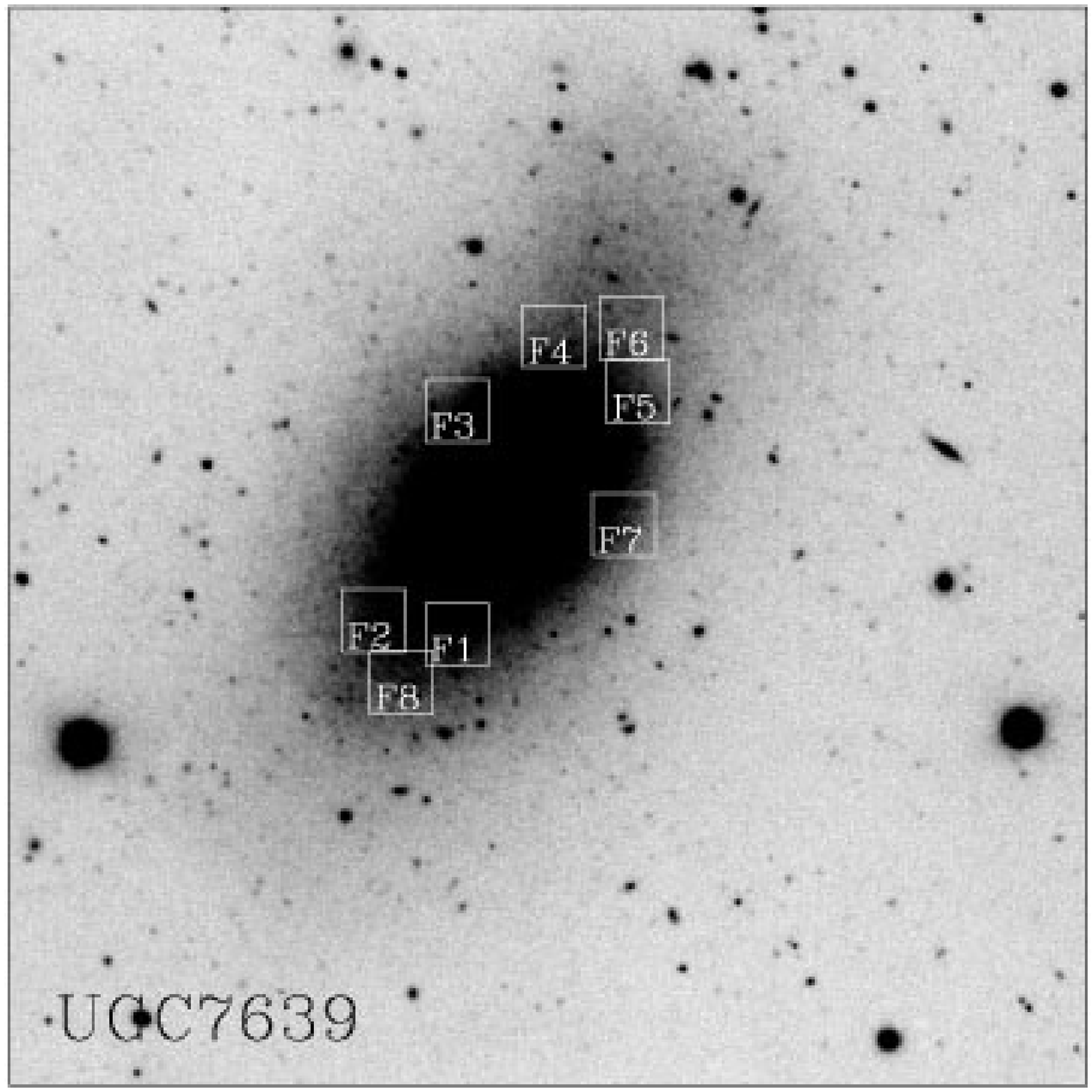}\\
\includegraphics[width=4.4cm]{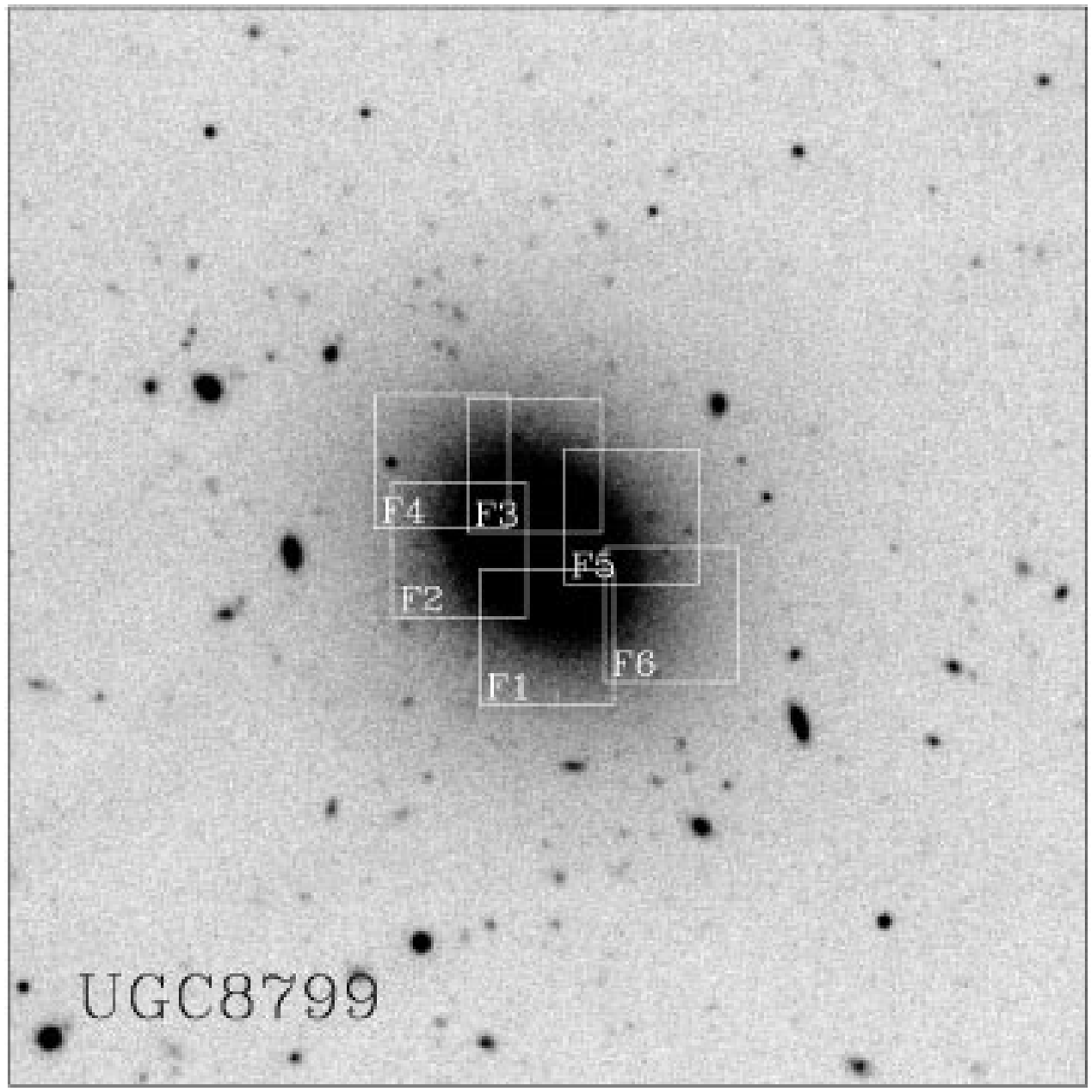}
\includegraphics[width=4.4cm]{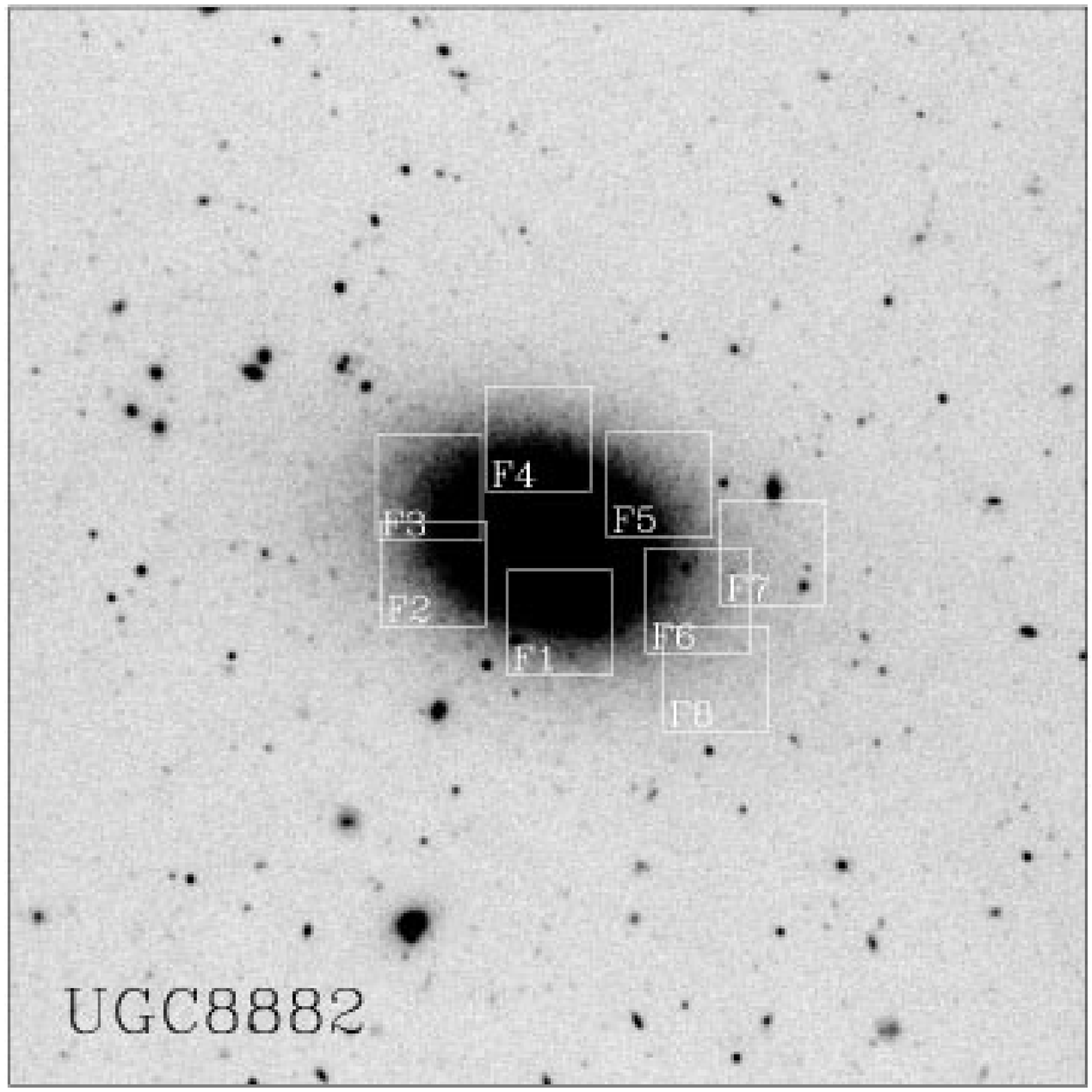}
\caption{Reduced $R$-band images of the ten sample galaxies 
with the boundaries of the analysed square SBF fields indicated. 
The field-of-view is $2.5 \times 2.5$ arcmin. North is up, East to the left.} 
\label{fields}
\end{figure*}

In Sect.~2, we describe the observations and data reduction. The SBF
analysis is presented in Sect.~3. Individual galaxies are discussed in 
Sect.~4. Finally, we present the summary and draw the conclusions of this 
work in Sect.~5.

\begin{table}
\caption{The sample of nearby early-type galaxies}
\label{tbl1}
\begin{tabular}{lllcc}
\hline \hline
         &           &         &    R.A.    &    Decl. \\
Name     & Location  & Type    & (J2000.0)  & (J2000.0) \\
\hline\small
UGC 1703 & NGC 784   & dE      & 02 12 55.8 & +32 48 51 \\
         & companion &         &            &           \\
KDG 61   & M81       & dE, N   & 09 57 03.1 & +68 35 31 \\
UGCA 200 & Field?    & dE, N   & 10 05 35.2 & $-$07 44 44 \\
UGC 5442 & M81       & dE, N   & 10 07 01.9 & +67 49 39 \\
UGC 5944 & Leo I     & dE/Irr  & 10 50 18.8 & +13 16 27 \\
NGC 4150 & CVn I?    & S0      & 12 10 33.6 & +30 24 06 \\
BTS 128  & Coma I    & dE      & 12 21 20.1 & +29 42 55 \\
UGC 7639 & CVn II    & dS0?    & 12 29 53.4 & +47 31 52 \\
UGC 8799 & Virgo I?  & dE      & 13 53 19.4 & +05 46 15 \\
UGC 8882 & M101      & dE, N   & 13 57 14.6 & +54 06 03 \\
\hline \hline
\end{tabular}
\end{table}

\begin{table}
\caption{Summary of observations}
\label{tbl2}
\begin{tabular}{lccccccc}
\hline \hline
         &             &  $t$          &     &      & FWHM \\
Name     &  Date       &  (sec)        & F   & $AM$ & ($\arcsec$)\\
\hline 
UGC 1703 & 20 Jan 2002 & 2$\times$600  & $B$ & 1.08 & 1.1 \\
         & 20 Jan 2002 & 6$\times$600  & $R$ & 1.03 & 0.9 \\
         & 04 Feb 2002 & 4$\times$600  & $B$ & 1.15 & 1.0 \\
         & 04 Feb 2002 & 1$\times$600  & $R$ & 1.09 & 0.8 \\
UGCA 200 & 03 Feb 2002 & 6$\times$600  & $B$ & 2.01 & 1.2 \\
         & 03 Feb 2002 & 6$\times$600  & $R$ & 1.38 & 0.8 \\
UGC 5944 & 04 Feb 2002 & 6$\times$600  & $B$ & 1.13 & 1.0 \\
         & 04 Feb 2002 & 6$\times$600  & $R$ & 1.05 & 0.9 \\
         & 25 Feb 2003 & 1$\times$600  & $B$ & 1.04 & 0.9 \\
         & 26 Feb 2003 & 1$\times$600  & $R$ & 1.04 & 1.0 \\
UGC 8882 & 25 Feb 2003 & 6$\times$600  & $R$ & 1.16 & 0.8 \\
         & 26 Feb 2003 & 5$\times$600  & $B$ & 1.12 & 0.9 \\
UGC 5442 & 26 Feb 2003 & 6$\times$600  & $R$ & 1.38 & 0.9 \\
         & 26 Feb 2003 & 5$\times$600  & $B$ & 1.31 & 0.9 \\
NGC 4150 & 26 Feb 2003 & 5$\times$600  & $B$ & 1.03 & 0.9 \\
         & 26 Feb 2003 & 10$\times$180 & $R$ & 1.16 & 0.8 \\
UGC 8799 & 26 Feb 2003 & 1$\times$600  & $B$ & 1.22 & 1.2 \\
         & 26 Feb 2003 & 3$\times$600  & $R$ & 1.16 & 1.0 \\
         & 27 Feb 2003 & 4$\times$600  & $B$ & 1.16 & 1.1 \\
         & 27 Feb 2003 & 3$\times$600  & $R$ & 1.10 & 1.0 \\
KDG 61   & 27 Feb 2003 & 5$\times$600  & $B$ & 1.49 & 1.1 \\
         & 27 Feb 2003 & 6$\times$600  & $R$ & 1.38 & 1.1 \\
BTS 128  & 27 Feb 2003 & 5$\times$600  & $B$ & 1.13 & 1.0 \\
         & 27 Feb 2003 & 6$\times$600  & $R$ & 1.03 & 0.8 \\
UGC 7639 & 27 Feb 2003 & 5$\times$600  & $B$ & 1.09 & 1.1 \\
         & 27 Feb 2003 & 6$\times$600  & $R$ & 1.06 & 0.9 \\
\hline\hline
\end{tabular}
\end{table}

\begin{table}
\caption[]{Photometric calibration coefficients}
\label{coeff}
\begin{tabular}{ccccr}
\hline \hline
Date        & F   & ZP               & $k$              & $c$ \hspace*{0.8cm}\\
\hline
03 Feb 2002 & $B$ & $25.40 \pm 0.03$ & $-0.22 \pm 0.02$ &  $0.002 \pm 0.042$ \\
            & $R$ & $25.22 \pm 0.02$ & $-0.08 \pm 0.01$ & $-0.029 \pm 0.061$ \\
25 Feb 2003 & $B$ & $26.06 \pm 0.12$ & $-0.48 \pm 0.08$ &  $0.040 \pm 0.011$ \\
            & $R$ & $25.73 \pm 0.09$ & $-0.31 \pm 0.06$ & $-0.031 \pm 0.019$ \\
26 Feb 2003 & $B$ & $25.77 \pm 0.04$ & $-0.26 \pm 0.02$ &  $0.042 \pm 0.013$ \\
            & $R$ & $25.47 \pm 0.04$ & $-0.10 \pm 0.02$ & $-0.030 \pm 0.006$ \\
27 Feb 2003 & $B$ & $25.68 \pm 0.09$ & $-0.24 \pm 0.05$ &  $0.046 \pm 0.022$ \\
            & $R$ & $25.66 \pm 0.12$ & $-0.16 \pm 0.07$ & $-0.028 \pm 0.021$ \\
\hline\hline
\end{tabular}
\end{table}

\section{Observations and Reductions}
CCD images were obtained at the 2.56 metre Nordic Optical Telescope on 
the nights of the 20 January 2002, 3-4 February 2002 and 25-27 February 
2003. We used the Andalucia Faint Object Spectrograph and Camera (ALFOSC), 
which is equipped with a 2048 $\times$ 2048 Loral/Lesser CCD detector with 
a pixel size of 15 $\mu$m and a plate scale of 0.188 arcsec, providing a 
field of view 6.4 arcmin on a side. The gain was set at 1 $e^- /$ADU. 
A series of six to ten images were taken in $B$ and $R$ passbands for 
each of the ten dwarf galaxies, along with bias frames, twilight flats 
and photometric standard star fields through the nights. The observing 
log is given in Table~\ref{tbl2}. The exposure time for individual science 
frames was 600 seconds, with the exception of NGC 4150 for which we used 
180 seconds in $R$ to avoid saturation of the central region of the galaxy. 
The seeing ranged from 0.8 to 1.6 arcsec and all six nights provided 
photometric conditions with the exception of the night of 4 February 2002, 
which was partly spectroscopic.

Image reduction was accomplished using routines within the 
IRAF\footnote{IRAF, Image Reduction and Analysis Facility, is 
distributed by the National Optical Astronomy Observatories, which is 
operated by the Association of Universities for Research in Astronomy, 
Inc., under contract with the National Science Foundation} programme. We 
removed the bias level from the images by using the bias frames and 
the overscan region of each image. Images were subsequently trimmed to 
2000 $\times$ 2000 pixels to remove non-essential data from the borders. 
Finally, each object image was divided by the corresponding median combined 
masterflat. Photometric calibration was achieved using the Landolt (1992) 
standard star fields, which were regularly observed during each night. 
Thus we determined the photometric zero point (ZP), atmospheric extinction 
coefficient ($k$) and colour term ($c$) for each passband and night. 
Analysis revealed slight variation in extinction coefficients throughout 
the observation period. The mean $k$ value was calculated for each passband 
and the corresponding values of ZP and $c$ were re-evaluated under 
this constraint. The results are summarised in Table~\ref{coeff}. Images 
taken during the nights of 20 Jan and 4 Feb 2002 (UGC1703, UGC5944) were 
calibrated with shallow images obtained on 26 Feb 2003.

$B$ and $R$ images of each galaxy were registered by matching the positions 
of typically 50 reference stars spread evenly over the image. The alignment 
was done on a pixel scale in order to avoid dividing galaxy flux in subpixel 
shifts. The resulting slight degradation of image quality is insignificant in 
relation to the seeing effects in the images. The sky background level was 
estimated by fitting a plane to selected star-free areas distributed uniformly 
over the CCD area but well away from the galaxy. The sky-subtracted images 
taken in the same passband were cleaned of cosmic rays and median-combined to 
increase the signal-to-noise. Finally, the resulting master images were flux 
calibrated.

\begin{table*}
\caption{Parameters of the SBF analysis. 
\label{parametertbl}}
\centering
\tiny
\begin{tabular}{ccccccccc}
\hline\hline
       &  size     & $m_1$  & $\overline{g}$ & $s$ & $P_0$  & $P_1$  & $S/N$ & $P_{\rm BG}/P_0$\\
 Name  & (pixels)  & (mag)  & (ADU)          & (ADU)         & (ADU s$^{-1}$ pixel$^{-1}$) & (ADU s$^{-1}$ pixel$^{-1}$)&    
&   \\
 (1)& (2) & (3) & (4)& (5) & (6) & (7) & (8)  & (9) \\
\hline
UGC 1703  F1 & 80 & 25.34 &   80.4 & 2314.5 & 0.189(0.009) & 0.011 & 15.9 &   0.00 \\ 
\dotfill  F2 & 80 &  &   66.8 &  & 0.171(0.007) & 0.014 & 11.5 &   0.00 \\ 
\dotfill  F3 & 80 &  &   70.6 &  & 0.207(0.007) & 0.012 & 16.1 &   0.00 \\ 
\dotfill  F4 & 80 &  &   66.1 &  & 0.165(0.007) & 0.013 & 11.9 &   0.01 \\ 
\dotfill  F5 & 80 &  &   99.1 &  & 0.183(0.007) & 0.009 & 18.5 &   0.00 \\ 
\hline
KDG 61    F1 & 100 & 25.47 &   68.2 & 1281.5 & 0.334(0.018) & 0.007 & 41.3 &   0.00 \\ 
\dotfill  F2 & 100 &  &   63.6 &  & 0.352(0.018) & 0.006 & 49.7 &   0.00 \\ 
\dotfill  F3 & 100 &  &   68.7 &  & 0.336(0.047) & 0.007 & 41.6 &   0.00 \\ 
\dotfill  F4 & 100 &  &   67.8 &  & 0.357(0.040) & 0.005 & 58.8 &   0.00 \\ 
\dotfill  F5 & 100 &  &   76.0 &  & 0.339(0.019) & 0.005 & 55.8 &   0.00 \\ 
\dotfill  F6 & 100 &  &   59.5 &  & 0.369(0.026) & 0.008 & 40.6 &   0.00 \\ 
\hline
UGCA 200  F1 & 80 & 25.17 &   63.6 & 2534.0 & 0.109(0.003) & 0.024 &  4.4 &   0.01 \\ 
\dotfill  F2 & 80 &  &   98.8 &  & 0.060(0.006) & 0.017 &  3.4 &   0.01 \\ 
\dotfill  F3 & 80 &  &  107.4 &  & 0.074(0.003) & 0.015 &  4.7 &   0.01 \\ 
\dotfill  F4 & 80 &  &  107.4 &  & 0.070(0.005) & 0.016 &  4.2 &   0.01 \\ 
\dotfill  F5 & 80 &  &   72.7 &  & 0.079(0.005) & 0.020 &  3.8 &   0.01 \\ 
\hline
UGC 5442\,F1 & 80 & 25.30 &  107.6 & 1441.1 & 0.243(0.047) & 0.005 & 42.0 &   0.00 \\ 
\dotfill  F2 & 80 &  &  105.2 &  & 0.221(0.015) & 0.005 & 38.1 &   0.00 \\ 
\dotfill  F3 & 80 &  &  111.7 &  & 0.227(0.018) & 0.004 & 47.4 &   0.00 \\ 
\dotfill  F4 & 80 &  &  113.6 &  & 0.241(0.015) & 0.004 & 50.3 &   0.00 \\ 
\dotfill  F5 & 80 &  &   87.2 &  & 0.241(0.013) & 0.005 & 41.6 &   0.00 \\ 
\dotfill  F6 &100 &  &   91.4 &  & 0.214(0.009) & 0.006 & 31.5 &   0.00 \\ 
\hline
UGC 5944  F1 & 70 & 25.34 &  172.7 & 2143.7 & 0.032(0.004) & 0.011 &  2.6 &   0.03 \\ 
\dotfill  F2 & 70 &  &  148.6 &  & 0.034(0.003) & 0.012 &  2.6 &   0.02 \\ 
\dotfill  F3 & 70 &  &  195.7 &  & 0.032(0.003) & 0.010 &  2.9 &   0.03 \\ 
\dotfill  F4 & 70 &  &  192.0 &  & 0.037(0.002) & 0.010 &  3.3 &   0.02 \\ 
\dotfill  F5 & 70 &  &  117.9 &  & 0.035(0.002) & 0.015 &  2.2 &   0.02 \\ 
\hline
NGC 4150\,F1 & 100 & 25.33 &   94.5 &  388.7 & 0.020(0.001) & 0.007 &  2.5 &   0.04 \\ 
\dotfill  F2 &100  &  & 141.3 &  & 0.018(0.001) & 0.006 &  2.5 &   0.05 \\ 
\dotfill  F3 &100  &  &   77.0 &  & 0.024(0.002) & 0.011 &  2.0 &   0.03 \\ 
\dotfill  F4 &100  &  &   97.7 &  & 0.021(0.002) & 0.009 &  2.1 &   0.04 \\ 
\dotfill  F5 &100  &  &  129.2 &  & 0.021(0.002) & 0.006 &  3.0 &   0.04 \\ 
\dotfill  F6 &100  &  &   40.7 &  & 0.033(0.003) & 0.014 &  2.2 &   0.02 \\ 
\hline
BTS 128   F1 & 80 & 25.52 &  174.1 & 1050.0 & 0.021(0.002) & 0.004 &  3.8 &   0.06 \\ 
\dotfill  F2 & 80 &  &  243.8 &  & 0.025(0.002) & 0.003 &  5.7 &   0.05 \\ 
\dotfill  F3 & 80 &  &   85.6 &  & 0.029(0.002) & 0.006 &  3.9 &   0.04 \\ 
\dotfill  F4 & 80 &  &  120.8 &  & 0.022(0.001) & 0.004 &  4.0 &   0.05 \\ 
\dotfill  F5 & 80 &  &  102.7 &  & 0.025(0.001) & 0.005 &  3.9 &   0.05 \\ 
\dotfill  F6 & 80 &  &  183.5 &  & 0.022(0.001) & 0.004 &  4.0 &   0.05 \\ 
\hline
UGC 7639  F1 & 70 & 25.52 &  220.8 & 1157.9 & 0.105(0.005) & 0.002 & 32.9 &   0.01 \\ 
\dotfill  F2 & 70 &  &  148.3 &  & 0.098(0.006) & 0.003 & 23.3 &   0.01 \\ 
\dotfill  F3 & 70 &  &  170.0 &  & 0.110(0.006) & 0.003 & 26.2 &   0.01 \\ 
\dotfill  F4 & 70 &  &  133.7 &  & 0.098(0.005) & 0.003 & 23.3 &   0.01 \\ 
\dotfill  F5 & 70 &  &  163.1 &  & 0.112(0.006) & 0.003 & 26.7 &   0.01 \\ 
\dotfill  F6 & 70 &  &  117.4 &  & 0.111(0.008) & 0.004 & 21.3 &   0.01 \\ 
\dotfill  F7 & 70 &  &  164.8 &  & 0.096(0.005) & 0.004 & 18.4 &   0.01 \\ 
\dotfill  F8 & 70 &  &  124.3 &  & 0.102(0.006) & 0.003 & 24.2 &   0.01 \\ 
\hline
UGC 8799  F1 & 100 & 25.32 &  184.7 & 1695.7 & 0.030(0.004) & 0.004 &  5.6 &   0.04 \\ 
\dotfill  F2 & 100 &  &  197.9 &  & 0.029(0.003) & 0.004 &  5.9 &   0.03 \\ 
\dotfill  F3 & 100 &  &  209.9 &  & 0.027(0.002) & 0.004 &  5.5 &   0.03 \\ 
\dotfill  F4 & 100 &  &  119.4 &  & 0.030(0.003) & 0.006 &  4.3 &   0.03 \\ 
\dotfill  F5 & 100 &  &  167.2 &  & 0.024(0.003) & 0.005 &  4.0 &   0.03 \\ 
\dotfill  F6 & 100 &  &   96.2 &  & 0.024(0.003) & 0.008 &  2.6 &   0.03 \\ 
\hline
UGC 8882\,F1 & 100 & 25.34 &  222.4 &  958.3 & 0.076(0.004) & 0.003 & 19.8 &   0.00 \\ 
\dotfill  F2 &  100 &  &  139.7 &  & 0.090(0.009) & 0.005 & 15.4 &   0.01 \\ 
\dotfill  F3 &  100 &  &  141.7 &  & 0.081(0.003) & 0.004 & 16.7 &   0.01 \\ 
\dotfill  F4 & 100  &  &  162.1 &  & 0.084(0.003) & 0.005 & 14.3 &   0.01 \\ 
\dotfill  F5 & 100  &  &  114.0 &  & 0.090(0.002) & 0.006 & 13.1 &   0.01 \\ 
\dotfill  F6 & 100  &  &   72.4 &  & 0.085(0.003) & 0.007 & 10.7 &   0.01 \\ 
\dotfill  F7 &  100 &  &   15.6 &  & 0.125(0.010) & 0.019 &  6.3 &   0.01 \\ 
\dotfill  F8 &  100 &  &   17.5 &  & 0.149(0.010) & 0.028 &  5.1 &   0.01 \\ 
\hline\hline
\end{tabular}
\end{table*}

\section{R-band SBF Analysis of Selected Early-Type Dwarf Galaxies in the 10 
Mpc Range}
We applied the Surface Brightness Fluctuation (SBF) method, developed to 
measure distances to dwarf elliptical galaxies by Jerjen et al.~(1998, 2000b) 
and calibrated in Jerjen et al.~(2001). The SBF analysis is done by carefully 
cleaning the galaxy images from any foreground stars, globular clusters, and 
background galaxies using procedures that follow the recipes of Jerjen et 
al.~(2000b; 2001). The cleaned galaxy image was then modelled using an 
isophote fitting routine written in IRAF that allows the centre, ellipticity, 
and position angle to vary. The best 2D-model was subtracted from the original 
master image and the residual image divided by the square root of the model 
for noise normalization. 

\begin{figure}
\centering
\includegraphics[width=6cm]{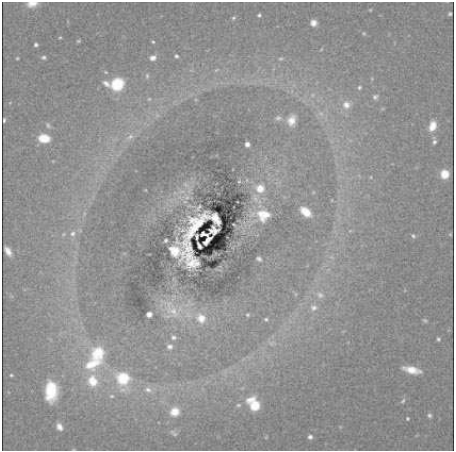}
\caption{After subtracting the best-fitting model of the global light 
distribution a well developed 4-armed spiral structure and traces of dust in 
the central region become visible in the residual image of NGC 4150. These 
findings confirm the morphological type of S0/Sa for this galaxy.} 
\label{ngc4150residual}
\end{figure}

Any non-radial irregularities in the light distribution of a galaxy such as 
the detected spiral arms and dust features in NGC 4150 remain visible in the 
residual image (see Fig.~\ref{ngc4150residual}). These parts of a galaxy were 
avoided in the SBF field selection. Only regions where the model follows 
closely the galaxy light distribution were used.

The largest possible number of slightly overlapping square subimages 
(hereafter SBF fields) were then defined for each galaxy. The size of the 
SBF fields was chosen between $70\times70$ and $100\times100$ pixels depending 
on the apparent diameter of the galaxy. Parts of a SBF field that were
contaminated by foreground stars or background galaxies were replaced with 
randomly selected patches from the fluctuation image, lying outside the field 
and in the same surface brightness range of the galaxy. The number of patched 
pixels was 3\% or less of the total SBF field area in all cases. In total we 
defined 61 SBF fields in our 10 sample galaxies. Their positions across the 
galaxy images are shown in Fig.~\ref{fields}. 

\begin{figure*}[h]
\centering
\includegraphics[height=6.5cm]{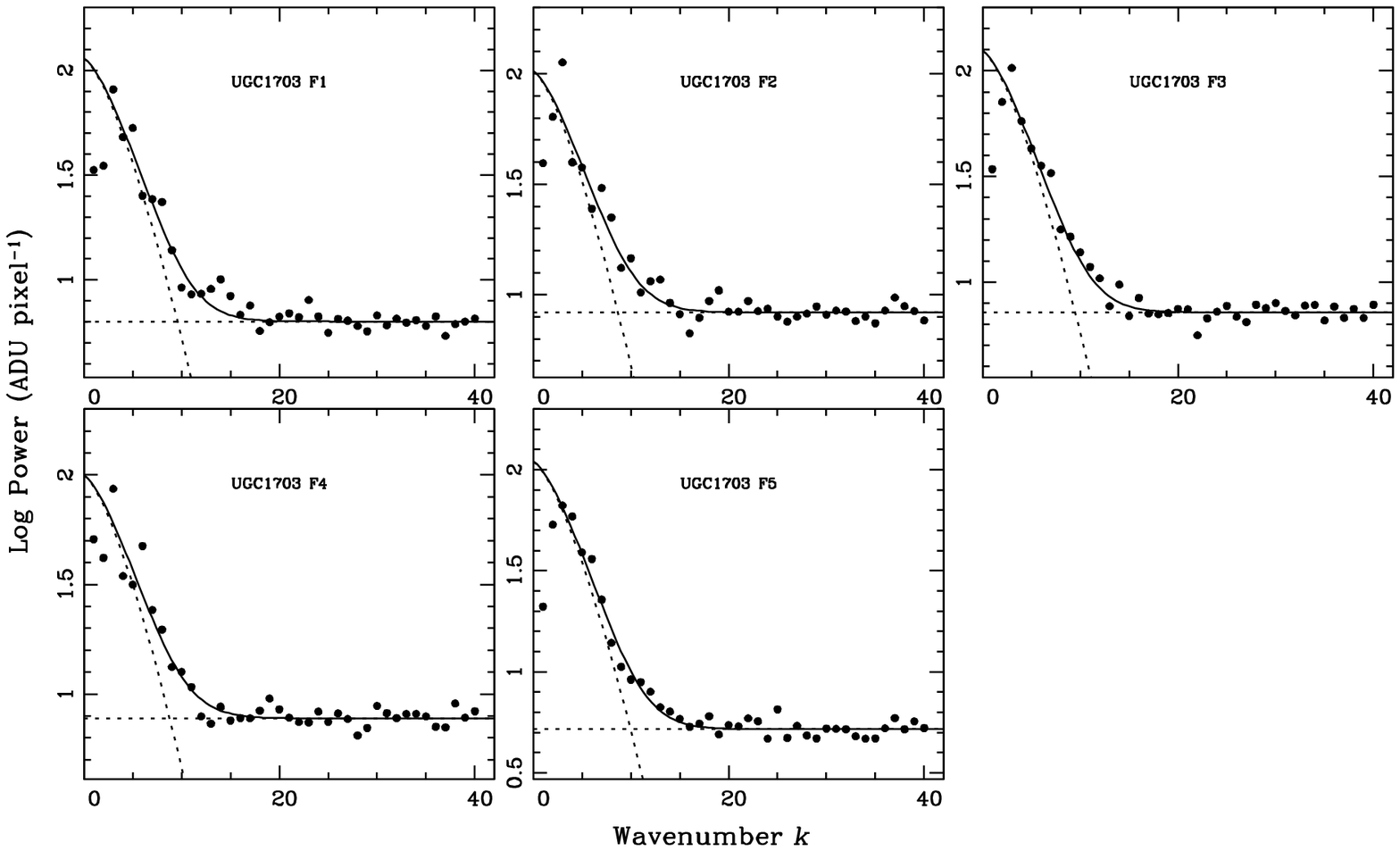}
\includegraphics[height=6.5cm]{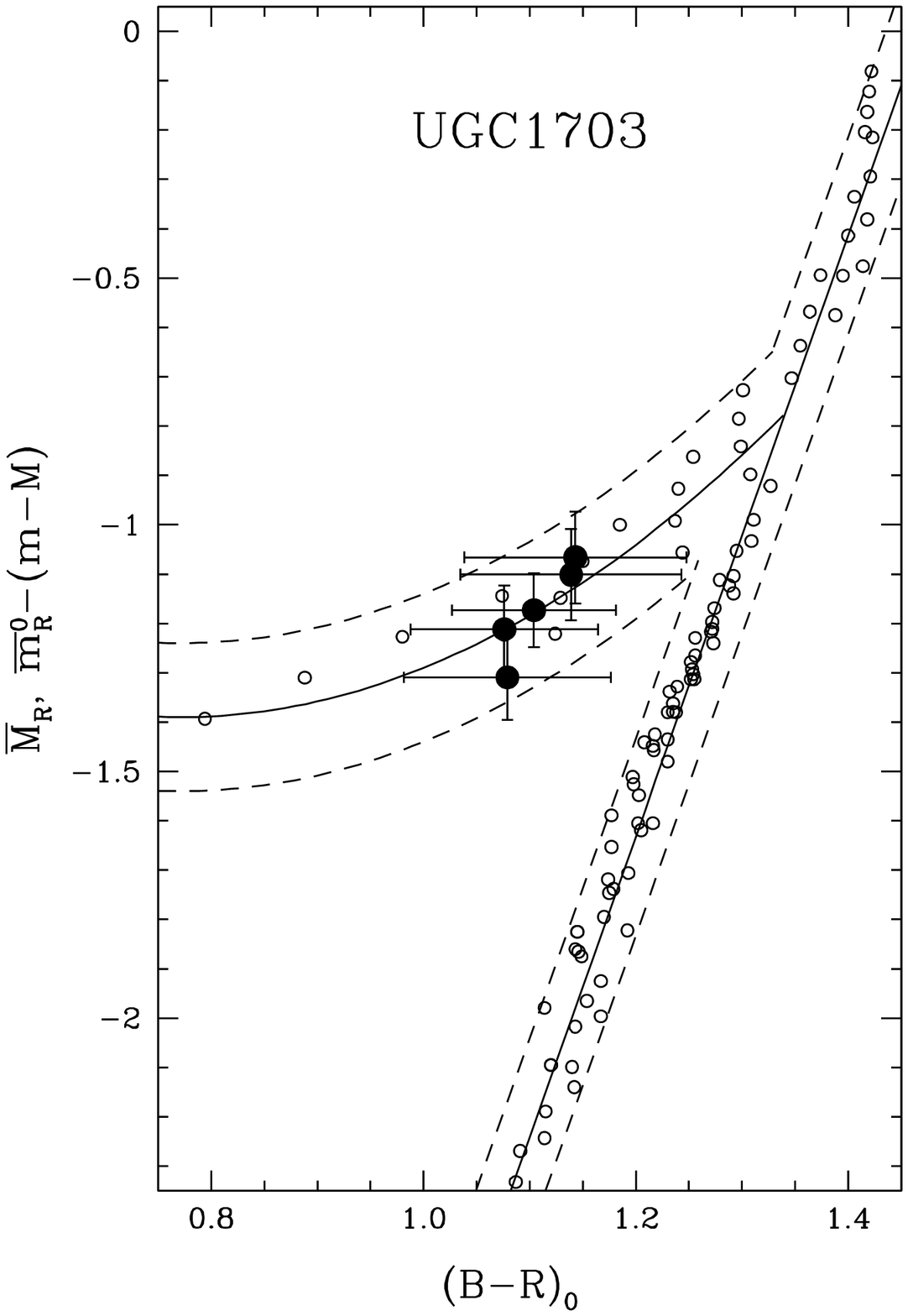}
\caption{
Left: Five fields were selected for the SBF analysis in UGC 1703. 
The signal-to-noise in the power spectra is generally high. The power spectra 
of the SBF fields (filled circles) are well fitted by the sum (solid line) of 
a scaled version of the power spectrum of the PSF and a constant (dashed 
lines). The wavenumbers 1--4 were not considered for the fit. 
Right: The fluctuation magnitudes $\overline{m}_R$ and $(B-R)$ colours were 
measured for the five fields to derive the distance of UGC 1703. A shift 
by 28.11\,mag yields the best fit of the data to the calibration diagram. 
}
\label{ugc1703}
\end{figure*}

\begin{figure*}[h]
\centering
\includegraphics[height=6.5cm]{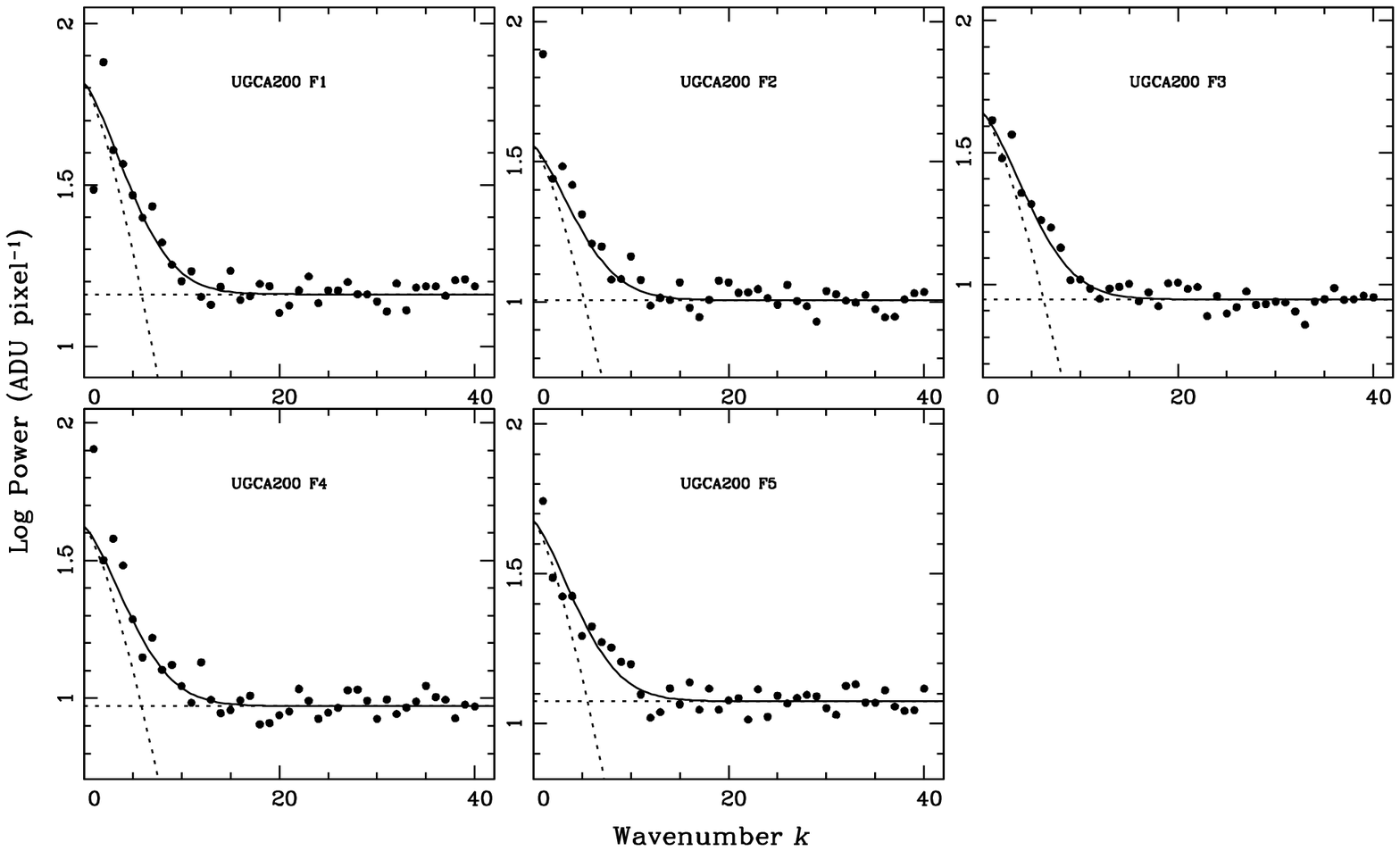}
\includegraphics[height=6.5cm]{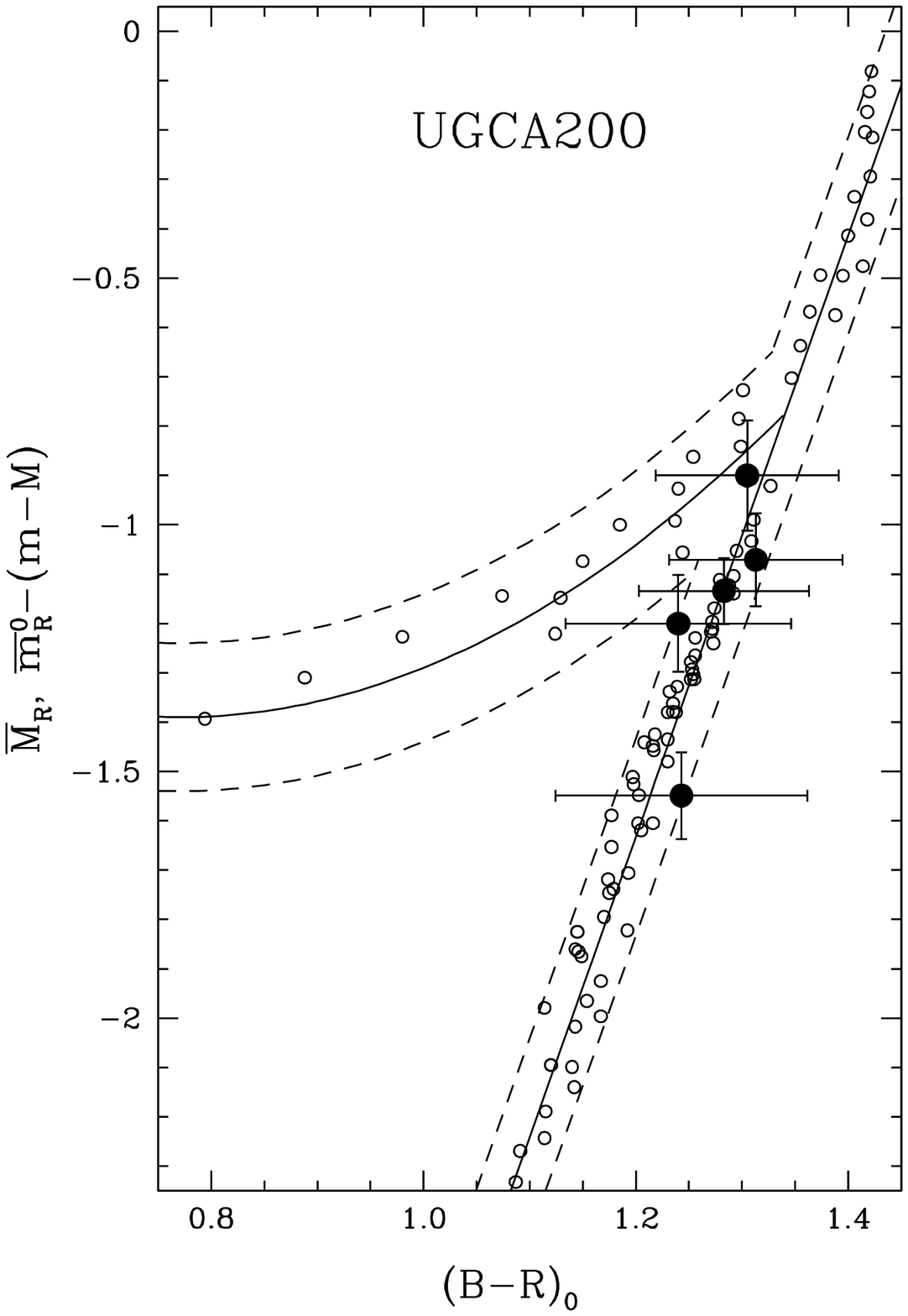}
\caption{
Left: Five fields were selected for the SBF analysis in UGCA 200. The power 
spectra of the SBF fields (filled circles) are well fitted by the sum (solid 
line) of a scaled version of the power spectrum of the PSF and a constant 
(dashed lines). The wavenumbers 1--4 were not considered for the fit. 
Right: The fluctuation magnitudes $\overline{m}_R$ and $(B-R)$ colours were 
measured for the five fields to derive the distance of UGCA 200. A shift by 
29.01\,mag yields the best fit of the data to the linear branch of the 
calibration diagram. 
}
\label{ugca200}
\end{figure*}

\begin{figure*}[h]
\centering
\includegraphics[height=6.5cm]{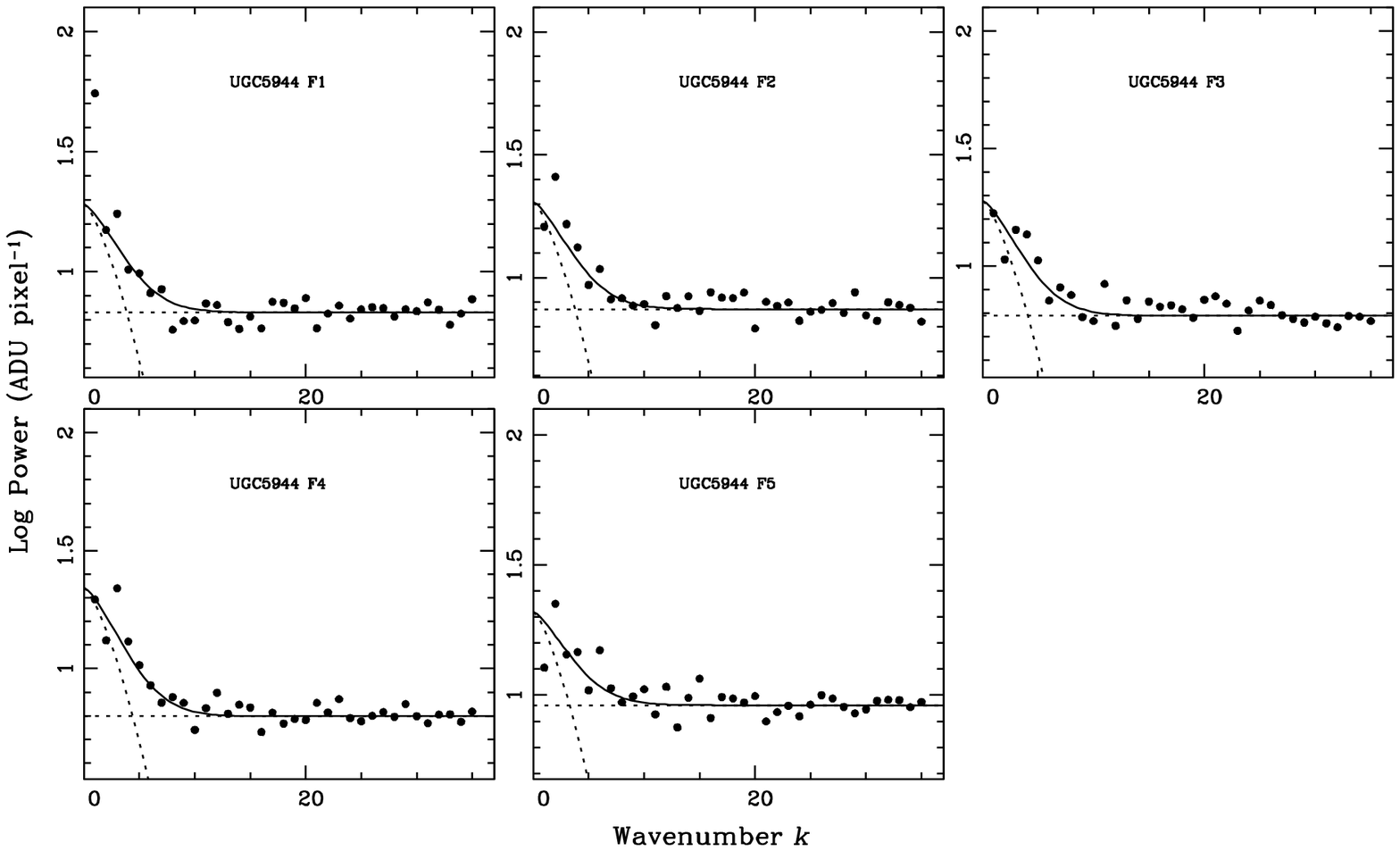}
\includegraphics[height=6.5cm]{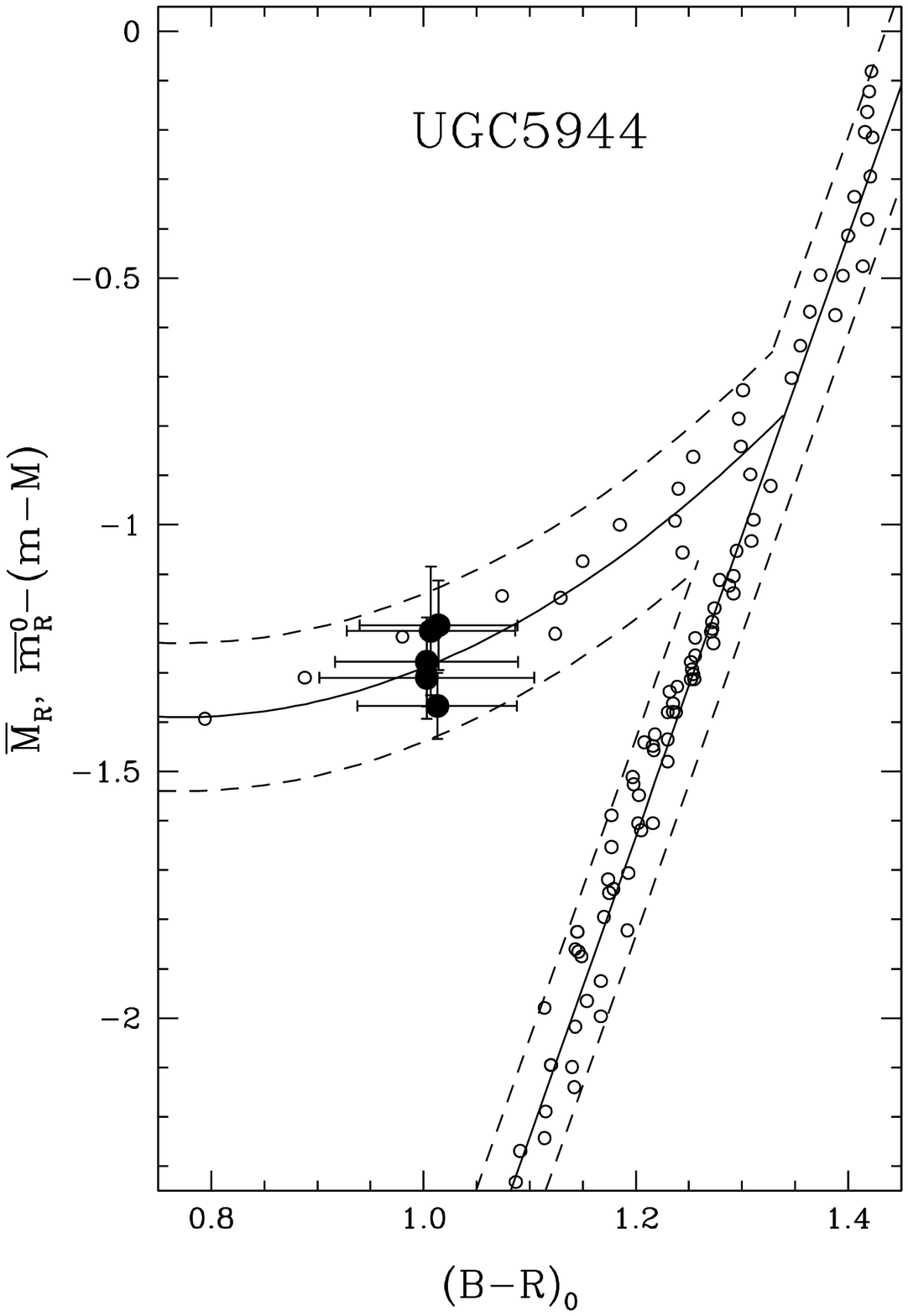}
\caption{
Left: Five fields were selected for the SBF analysis in UGC 5944. The 
signal-to-noise in the power spectra is not as high as measured in the other 
galaxies due to the larger distance of the galaxy at 11.1\,Mpc. Nevertheless, 
the power spectra of the SBF fields (filled circles) are well fitted by the 
sum (solid line) of a scaled version of the power spectrum of the PSF and a 
constant (dashed lines). The wavenumbers 1--4 were not considered for the fit.
Right: Five fields were analysed to derive the distance of UGC 5944. A shift 
by 30.22\,mag yields the best fit of the data to the calibration diagram. 
}
\label{ugc5944}
\end{figure*}

\begin{figure*}[h]
\centering
\includegraphics[height=6.5cm]{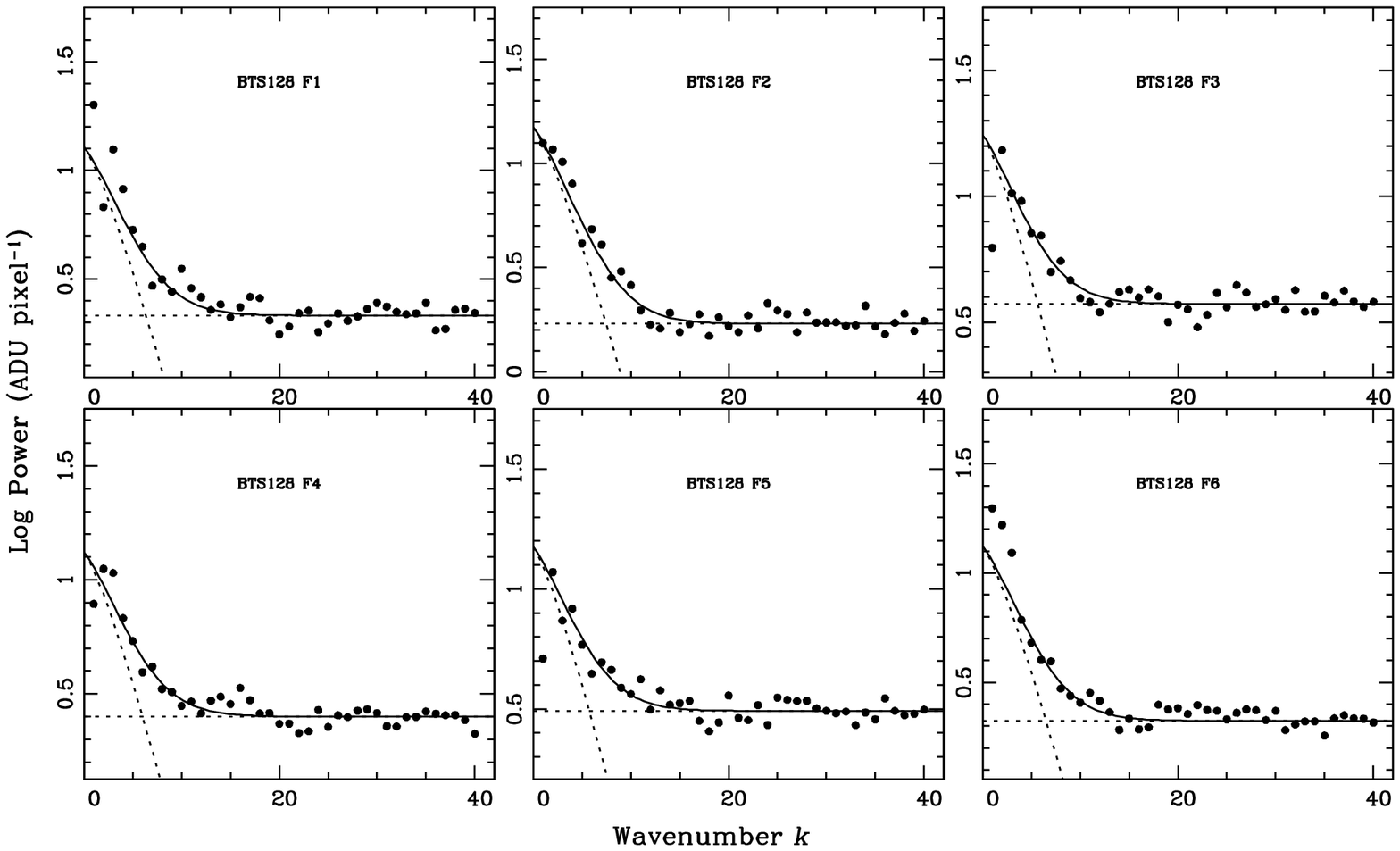}
\includegraphics[height=6.5cm]{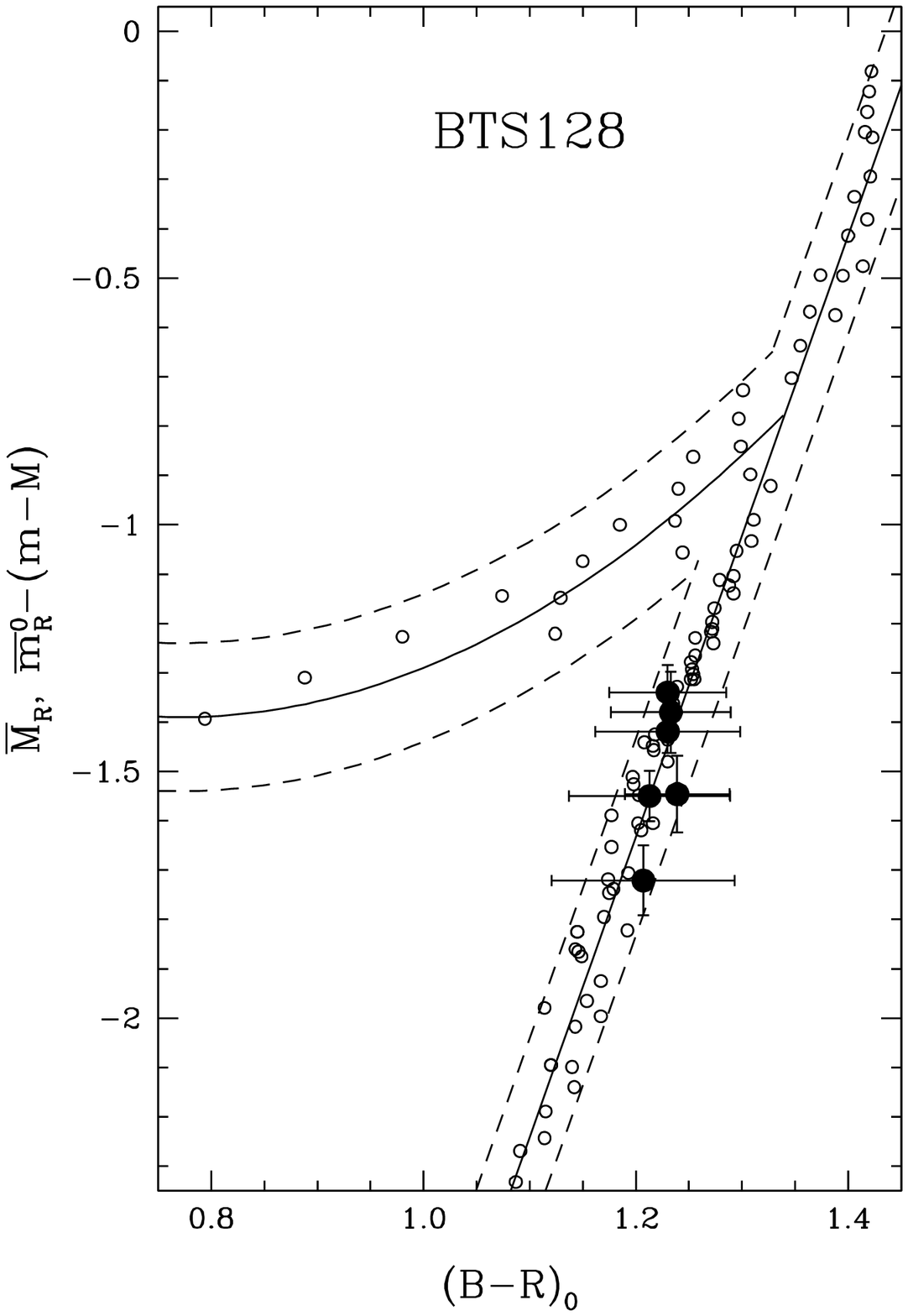}
\caption{
Left: Six fields were selected for the SBF analysis in BTS 128. The power 
spectra of the SBF fields (filled circles) are well fitted by the sum (solid 
line) of a scaled version of the power spectrum of the PSF and a constant 
(dashed lines). The wavenumbers 1--4 were not considered for the fit. 
Right: Six fields were analysed to derive the distance of BTS 128. A shift 
by 31.02\,mag yields the best fit of the data to the calibration diagram. 
} 
\label{bts128}
\end{figure*}

All SBF fields were Fourier transformed and the azimuthally averaged power 
spectra calculated. From isolated bright stars on the galaxy master image we 
determined the point spread function (PSF) profile. We then fitted a linear 
combination of the flux normalized and exposure time weighted PSF power 
spectrum and a constant at the observed galaxy power spectrum 
$\mbox{PS}(k)=P_0\,\mbox{PS}_{\mbox{star}}(k) + P_1$, demanding a least 
squares minimization. Data points at low spatial frequencies ($k\le5$) were 
omitted as they are likely affected by imperfect galaxy model subtraction. 
Figs.~\ref{ugc1703} to \ref{bts128} and \ref{ugc7639} to \ref{ugc8882} show 
the power spectrum of each SBF field with the best fitting analytic function 
indicated as solid lines. Table~\ref{parametertbl} summarizes the quantities 
measured in the SBF analysis: 
Col.~1 -- SBF field number and galaxy name, 
Col.~2 -- pixel size of the SBF field, 
Col.~3 -- magnitude $m_1$ of a star yielding 1 ADU per second on the CCD, 
Col.~4 -- mean galaxy surface brightness within the SBF field in ADU, 
Col.~5 -- sky brightness in ADU, 
Col.~6 -- exposure time normalized amplitude $P_0$ of the best least 
squares fit at wavenumber $k=0$ with fitting error in brackets, 
Col.~7 -- the scale-free white noise component $P_1$ in the power spectrum, 
indicating the ratio of sky to mean galaxy surface brightness within the 
SBF field. 

To estimate the fraction in $P_0$ (Col.~6) from unresolved distant background 
galaxies fainter than the cutoff magnitude $m_c=24.7$\,$R$\,mag, we made 
use of a formula that was given in Jensen et al.~(1998) and adjusted for the 
$R$-band by Jerjen et al.~(2001):   

$$P_{\rm BG}={{p^2} \over 
{(0.8-\gamma)\ln 10}}10^{0.8(m_1-m_c)-\gamma(29.38+R-K-m_c)},$$

\noindent where $p$ is the CCD pixel size in arcsec and $\gamma=0.3$ 
the slope of the power-law number distribution for background galaxies 
in the $K$-band (Cowie et al.~1994). Assuming a typical galaxy colour of 
$(R-K)=2.25$ (de Jong 1996) we computed $P_{\rm BG}$ and determined the 
signal-to-noise S/N$=(P_0-P_{\rm BG})/(P_1+P_{\rm BG})$ as well as the 
relative contribution to the signal $P_{\rm BG}/P_0$ for each individual SBF 
field. Both numbers are listed in Col.~8 and Col.~9 of 
Table~\ref{parametertbl}. The contribution from unresolved background 
galaxies was minimal at the 0--6 per cent level in our SBF fields. 

Another potentially significant source of unwanted fluctuations is a rich 
globular cluster (GC) system in a target galaxy. While this is an important 
issue for luminous giant ellipticals the expected number of GCs in our dwarf 
ellipticals is quite small. For instance, the net number of globular cluster 
candidates for UGC 5944 in the Leo Group is $3.5\pm3.9$ (Miller et al.~1998). 
The GC frequency ($S_N$)--luminosity relation 
for dE,Ns studied in the Fornax and Virgo clusters (Miller et al.~1998) 
predicts $\approx 18$ GCs (assuming $S_N=2$) for our brightest dwarf 
NGC 4150 ($M_V \approx -17.4$) and $\approx 0$ GCs (assuming $S_N=10$) for 
the faintest dwarf UGC 5442 ($M_V \approx -10$). All GCs would be brighter 
than our cutoff luminosity and thus be excised during the image cleaning 
process. Therefore, no further corrections were applied to the measured SBF 
power. 

Finally, we calculated the stellar fluctuation magnitude $\overline{m}_R$ 
with the formula $\overline{m}_R=m_1-2.5\log(P_0-P_{\rm BG})$ and 
measured the $(B-R)$ colour for each SBF field from the cleaned 
$B$ and $R$ galaxy master images. Both quantities were corrected for 
foreground extinction using the IRAS/DIRBE maps of dust IR emission 
(Schlegel et al.~1998). The results are listed in  
Cols.~3 and 4 of Table~\ref{flucmagcol}. 

The power spectrum fitting error is between 3 and 15\%. Other 
sources of minor errors are the PSF normalization ($\sim$2\%), 
the shape variation of the stellar PSF over the CCD area (1--2\%) 
and the uncertainty in the photometric calibration ($0.04$\,mag in $B$, 
0.03\,mag in $R$). If we further adopt a 16\% error for the foreground 
extinction (Schlegel et al.~1998), the formal combined error for a 
single $\overline{m}_R^0$ measurement is between 0.05 and 0.20\,mag 
(Col.~3). The error associated with the local colour (Col.~4) 
has been obtained through the usual error propagation 
formula from the uncertainties in the sky level determination, 
the photometry zero points, and Galactic extinction.

\begin{table}
\caption{Fluctuation magnitudes and local colours for each SBF field 
in the dwarfs after correction for Galactic extinction.
\label{flucmagcol}}
\centering
\tiny
\begin{tabular}{cccc}
\hline\hline
\small
       & A$_R$ & $\overline{m}_R^0$&$(B-R)_0$ \\
 Name  & (mag) & (mag)                                  & (mag)    \\
 (1)& (2) & (3) & (4)\\
\hline 
NGC 1703 F1 &  $ 0.26 \pm  0.04$ & $26.89 \pm 0.089$ & $ 1.08 \pm  0.10$ \\ 
\dotfill F2 &                    & $27.00 \pm 0.092$ & $ 1.14 \pm  0.12$ \\ 
\dotfill F3 &                    & $26.79 \pm 0.086$ & $ 1.08 \pm  0.11$ \\ 
\dotfill F4 &                    & $27.04 \pm 0.093$ & $ 1.14 \pm  0.12$ \\ 
\dotfill F5 &                    & $26.93 \pm 0.076$ & $ 1.10 \pm  0.09$ \\ 
\hline
KDG 61     F1 &  $ 0.19 \pm  0.03$ & $26.47 \pm 0.077$ & $ 1.09 \pm  0.12$ \\ 
\dotfill F2 &                    & $26.41 \pm 0.076$ & $ 1.11 \pm  0.13$ \\ 
\dotfill F3 &                    & $26.46 \pm 0.149$ & $ 1.10 \pm  0.12$ \\ 
\dotfill F4 &                    & $26.40 \pm 0.125$ & $ 1.09 \pm  0.12$ \\ 
\dotfill F5 &                    & $26.45 \pm 0.077$ & $ 1.10 \pm  0.11$ \\ 
\dotfill F6 &                    & $26.36 \pm 0.091$ & $ 1.11 \pm  0.14$ \\ 
\hline
UGCA 200 F1 &  $ 0.13 \pm  0.02$ & $27.45 \pm 0.087$ & $ 1.24 \pm  0.14$ \\ 
\dotfill F2 &                    & $28.11 \pm 0.113$ & $ 1.31 \pm  0.10$ \\ 
\dotfill F3 &                    & $27.87 \pm 0.069$ & $ 1.28 \pm  0.09$ \\ 
\dotfill F4 &                    & $27.94 \pm 0.095$ & $ 1.31 \pm  0.09$ \\ 
\dotfill F5 &                    & $27.80 \pm 0.097$ & $ 1.24 \pm  0.13$ \\ 
\hline
UGC 5442 \, F1 &  $ 0.14 \pm  0.02$ & $26.70 \pm 0.197$ & $ 1.24 \pm  0.08$ \\ 
\dotfill F2 &                    & $26.80 \pm 0.080$ & $ 1.22 \pm  0.08$ \\ 
\dotfill F3 &                    & $26.77 \pm 0.092$ & $ 1.24 \pm  0.08$ \\ 
\dotfill F4 &                    & $26.71 \pm 0.077$ & $ 1.22 \pm  0.08$ \\ 
\dotfill F5 &                    & $26.71 \pm 0.073$ & $ 1.23 \pm  0.10$ \\ 
\dotfill F6 &                    & $26.84 \pm 0.062$ & $ 1.23 \pm  0.09$ \\ 
\hline
NGC 5944  F1 &  $ 0.08 \pm  0.01$ & $29.02 \pm 0.131$ & $ 1.01 \pm  0.07$ \\ 
\dotfill F2 &                    & $28.96 \pm 0.092$ & $ 1.00 \pm  0.08$ \\ 
\dotfill F3 &                    & $29.03 \pm 0.093$ & $ 1.01 \pm  0.06$ \\ 
\dotfill F4 &                    & $28.87 \pm 0.070$ & $ 1.01 \pm  0.06$ \\ 
\dotfill F5 &                    & $28.93 \pm 0.085$ & $ 1.00 \pm  0.09$ \\ 
\hline
NGC 4150 \, F1 &  $ 0.05 \pm  0.01$ & $29.54 \pm 0.063$ & $ 1.24 \pm  0.07$ \\ 
\dotfill F2 &                    & $29.64 \pm 0.054$ & $ 1.26 \pm  0.05$ \\ 
\dotfill F3 &                    & $29.35 \pm 0.084$ & $ 1.23 \pm  0.08$ \\ 
\dotfill F4 &                    & $29.48 \pm 0.079$ & $ 1.24 \pm  0.06$ \\ 
\dotfill F5 &                    & $29.47 \pm 0.081$ & $ 1.25 \pm  0.05$ \\ 
\dotfill F6 &                    & $28.98 \pm 0.099$ & $ 1.17 \pm  0.13$ \\ 
\hline 
BTS 128    F1 &  $ 0.07 \pm  0.01$ & $29.68 \pm 0.085$ & $ 1.23 \pm  0.06$ \\ 
\dotfill F2 &                    & $29.51 \pm 0.080$ & $ 1.24 \pm  0.05$ \\ 
\dotfill F3 &                    & $29.33 \pm 0.073$ & $ 1.21 \pm  0.10$ \\ 
\dotfill F4 &                    & $29.64 \pm 0.047$ & $ 1.23 \pm  0.07$ \\ 
\dotfill F5 &                    & $29.50 \pm 0.055$ & $ 1.21 \pm  0.08$ \\ 
\dotfill F6 &                    & $29.64 \pm 0.065$ & $ 1.23 \pm  0.05$ \\ 
\hline
UGC 7639 F1 &  $ 0.03 \pm  0.00$ & $27.95 \pm 0.057$ & $ 0.90 \pm  0.04$ \\ 
\dotfill F2 &                    & $28.02 \pm 0.069$ & $ 0.96 \pm  0.06$ \\ 
\dotfill F3 &                    & $27.89 \pm 0.061$ & $ 0.93 \pm  0.05$ \\ 
\dotfill F4 &                    & $28.03 \pm 0.063$ & $ 0.95 \pm  0.06$ \\ 
\dotfill F5 &                    & $27.88 \pm 0.065$ & $ 0.95 \pm  0.05$ \\ 
\dotfill F6 &                    & $27.88 \pm 0.077$ & $ 0.97 \pm  0.07$ \\ 
\dotfill F7 &                    & $28.05 \pm 0.058$ & $ 0.96 \pm  0.05$ \\ 
\dotfill F8 &                    & $27.98 \pm 0.069$ & $ 0.97 \pm  0.06$ \\ 
\hline
UGC 8799 F1 &  $ 0.07 \pm  0.01$ & $29.08 \pm 0.123$ & $ 1.24 \pm  0.05$ \\ 
\dotfill F2 &                    & $29.13 \pm 0.105$ & $ 1.24 \pm  0.05$ \\ 
\dotfill F3 &                    & $29.18 \pm 0.085$ & $ 1.25 \pm  0.05$ \\ 
\dotfill F4 &                    & $29.10 \pm 0.099$ & $ 1.23 \pm  0.08$ \\ 
\dotfill F5 &                    & $29.31 \pm 0.122$ & $ 1.23 \pm  0.06$ \\ 
\dotfill F6 &                    & $29.35 \pm 0.147$ & $ 1.22 \pm  0.09$ \\ 
\hline
UGC 8882 \, F1 &  $ 0.02 \pm  0.00$ & $28.13 \pm 0.058$ & $ 1.22 \pm  0.05$ \\ 
\dotfill F2 &                    & $27.94 \pm 0.108$ & $ 1.21 \pm  0.06$ \\ 
\dotfill F3 &                    & $28.05 \pm 0.047$ & $ 1.21 \pm  0.06$ \\ 
\dotfill F4 &                    & $28.01 \pm 0.050$ & $ 1.21 \pm  0.06$ \\ 
\dotfill F5 &                    & $27.95 \pm 0.043$ & $ 1.20 \pm  0.07$ \\ 
\dotfill F6 &                    & $28.01 \pm 0.055$ & $ 1.18 \pm  0.11$ \\ 
\dotfill F7 &                    & $27.59 \pm 0.146$ & $ 1.07 \pm  0.21$ \\ 
\dotfill F8 &                    & $27.39 \pm 0.125$ & $ 1.09 \pm  0.18$ \\ 
\hline
\hline                                               
\end{tabular}
\end{table}

\section{Discussion}
\subsection*{UGC 1703}
Neither a velocity nor a distance were known for UGC 1703 to date. This dwarf 
galaxy is closest in projection (3 degrees) to the spiral galaxy NGC 925 for 
which a Cepheid distance of $9.3 \pm 0.7$\,Mpc (Silbermann et al.~1996) and a 
velocity of $v_{\odot} = 553$\,km\,s$^{-1}$ were reported. However, our SBF 
distance for UGC 1703 of $4.2 \pm 0.3$\,Mpc indicates a much shorter distance 
from the Milky Way and thus UGC 1703 seems to be only close to NGC 925 in 
projection. The next nearest major galaxy to UGC 1703 is the SBdm spiral NGC 
784 with an angular separation of 4.7 degrees. NGC 784 has a velocity of 
$v_{\odot} = 198$\,km\,s$^{-1}$ and a reported distance of 5\,Mpc 
(Drozdovsky \& Karachentsev 2000). These two results confirm that UGC 1703 is 
spatially close to NGC 784 with a projected linear distance of only 
$\sim$0.4\,Mpc at an adopted distance of 4.5\,Mpc. Another more qualitative 
evidence for the short distance of UGC 1703 comes from our deep $R$-band 
image (Fig.\ref{fields}) which shows semi-resolved stars in appearance similar 
to images of dwarf ellipticals in the Cen\,A Group observed with a 2.3m 
ground-based telescope (Jerjen et al.~2000a).

\subsection*{KDG 61 ([KK98] 81)}
\begin{figure*}[h]
\centering
\includegraphics[height=6.5cm]{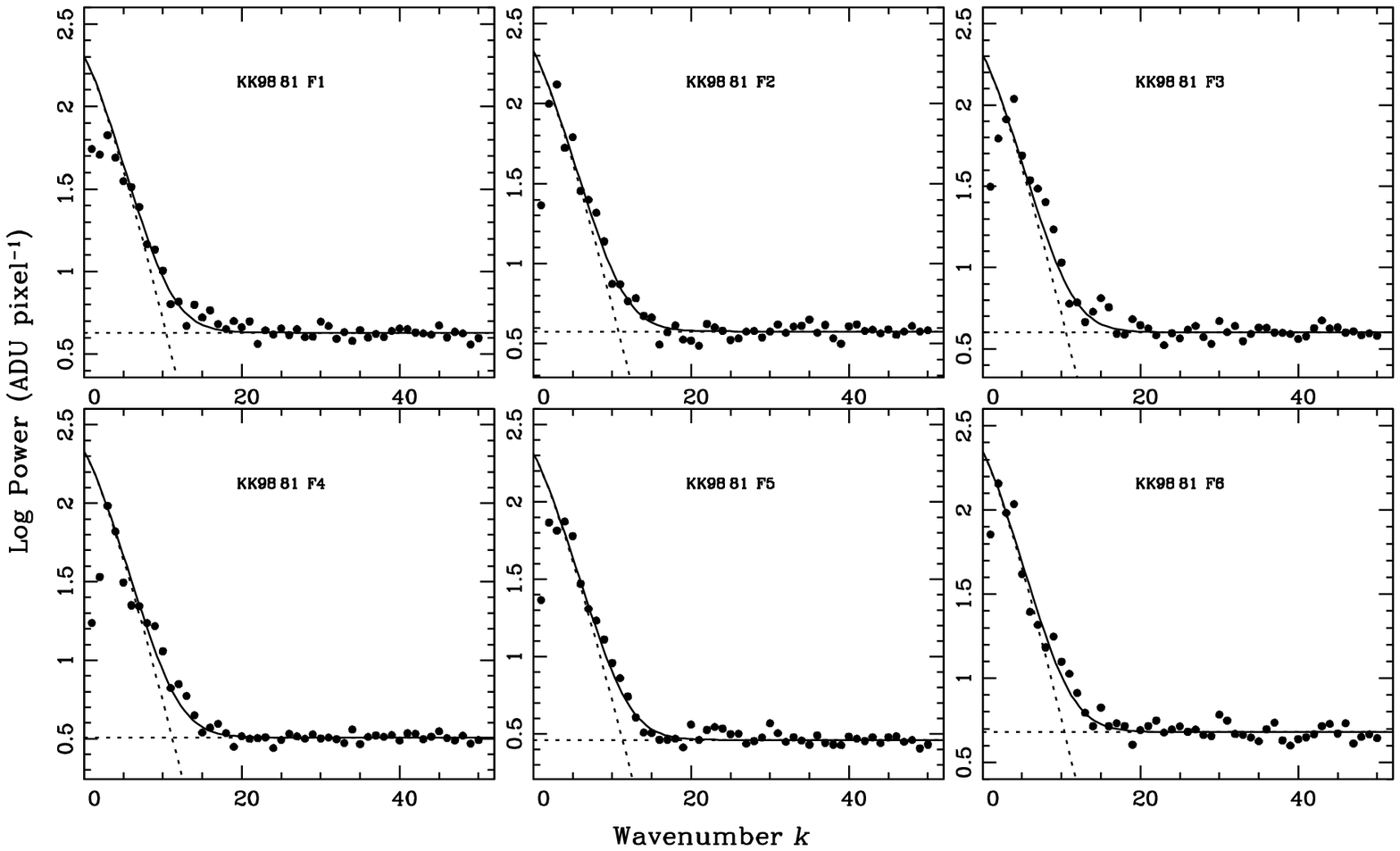}
\includegraphics[height=6.5cm]{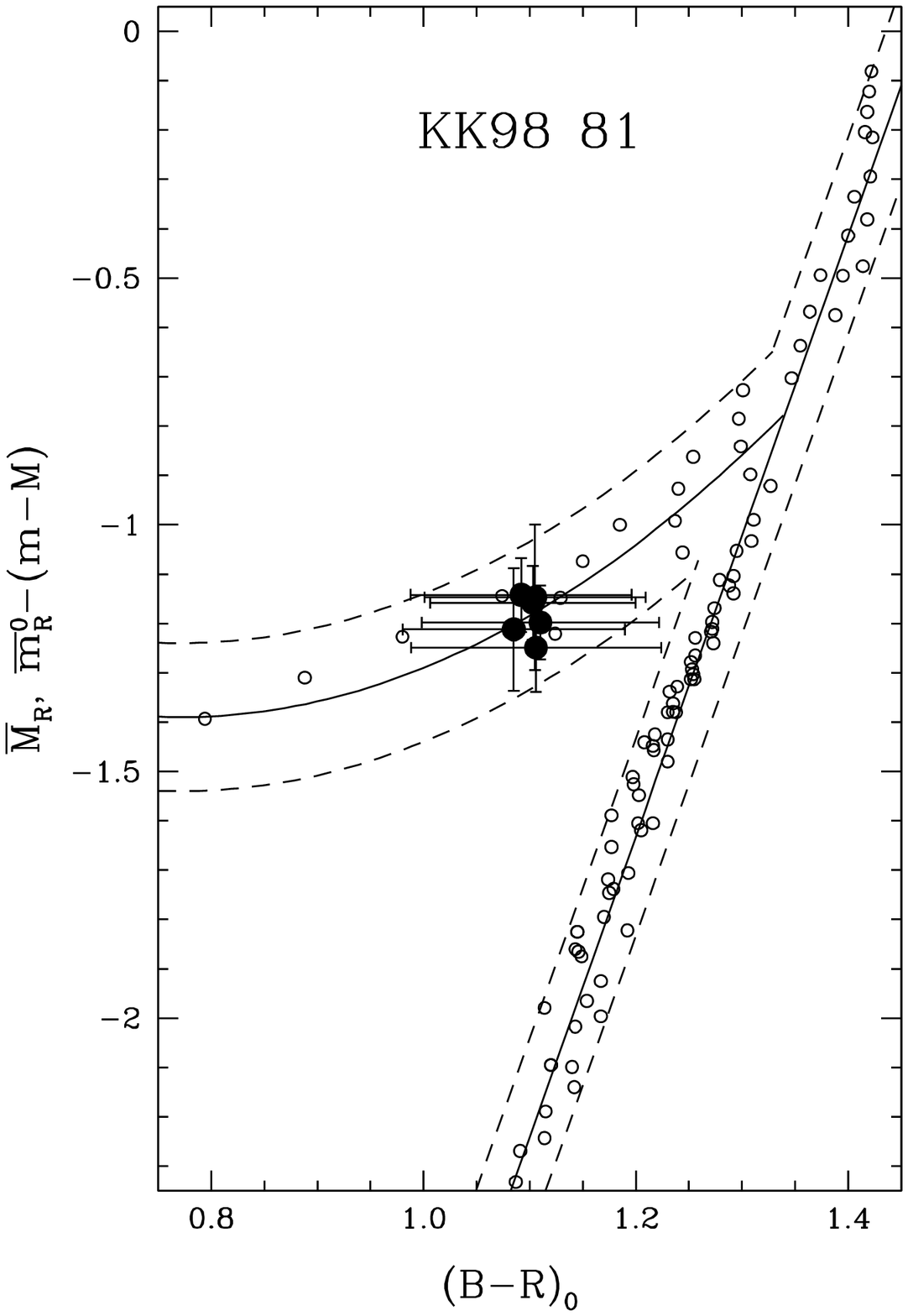}
\caption{
Left: Six fields were selected for the SBF analysis in KDG 61. The power 
spectra of the SBF fields (filled circles) are well fitted by the sum (solid 
line) of a scaled version of the power spectrum of the PSF and a constant 
(dashed lines). The wavenumbers 1--4 were not considered for the fit. 
Right: The fluctuation magnitudes $\overline{m}_R$ and $(B-R)$ colours were 
measured for the six fields to derive the distance of KDG 61. A shift 
by 27.80\,mag yields the best fit of the data to the calibration diagram. 
}
\label{kdg61}
\end{figure*}

KDG 61 ([KK98] 81) is a member of the M81 group. Karachentsev et al.~(2000) 
reported a distance modulus of $(m-M) = 27.78 \pm 0.15$\,mag from the 
measurement of the magnitude of the red giant branch tip and the galaxy has 
a heliocentric velocity of $v_{\odot} = -135 \pm 30$\,km\,s$^{-1}$ 
(Johnson et al.~1997). We analysed six independent SBF fields across the 
galaxy's surface (Fig.~\ref{kdg61}, left panels). The derived mean distance 
modulus of $(m-M)_{\rm{SBF}} = 27.80 \pm 0.20$\,mag (see Fig.~\ref{kdg61}, 
right panel) is in good agreement with the TRGB result. Due to the small 
colour range covered by the SBF fields (see Table \ref{flucmagcol}), another 
SBF distance is technically possible by moving the data points onto the linear 
branch of the calibration curve. However, that alternative distance modulus of 
$(m-M)_{\rm{SBF}} = 30.0 \pm 0.3$\,mag is highly inconsistent with the 
independent TRGB result and thus can be ruled out. 

\subsection*{UGCA 200}
Neither a velocity nor a distance were known for UGCA 200 to date. This dE,N 
was previously photometrically studied by Parodi et al. (2002). The authors 
reported an outward colour gradient getting redder and an integrated colour of 
$(B-R)_0=1.38$. Our five SBF fields were selected from the inner region of the 
dwarf and thus are slightly bluer, in the range $1.22<(B-R)_0<1.31$. 
Nevertheless, the SBF fields of UGCA 200 are the reddest in our sample. The 
observed correlation between the derived parameters $\overline{m}_R$ and 
$(B-R)$ colour allowed an unambiguous measurement of the distance. A shift by 
29.01\,mag yields the best fit to the linear branch of the calibration 
diagram. UGCA 200 was assumed to be a faint companion of the S0 galaxy NGC 
3115 which has a velocity of $v_{\odot} = 720$\,km\,s$^{-1}$ and an accurate 
SBF distance of $9.7 \pm 0.1$\,Mpc (Tonry et al. 2001). However, our SBF 
distance for UGCA 200 is significantly shorter with $6.3 \pm 0.8$\,Mpc and 
thus suggests that this dwarf is actually in the foreground of NGC 3115.

\subsection*{UGC 5442 (KDG 64)}
\begin{figure*}[h]
\centering
\includegraphics[height=6.5cm]{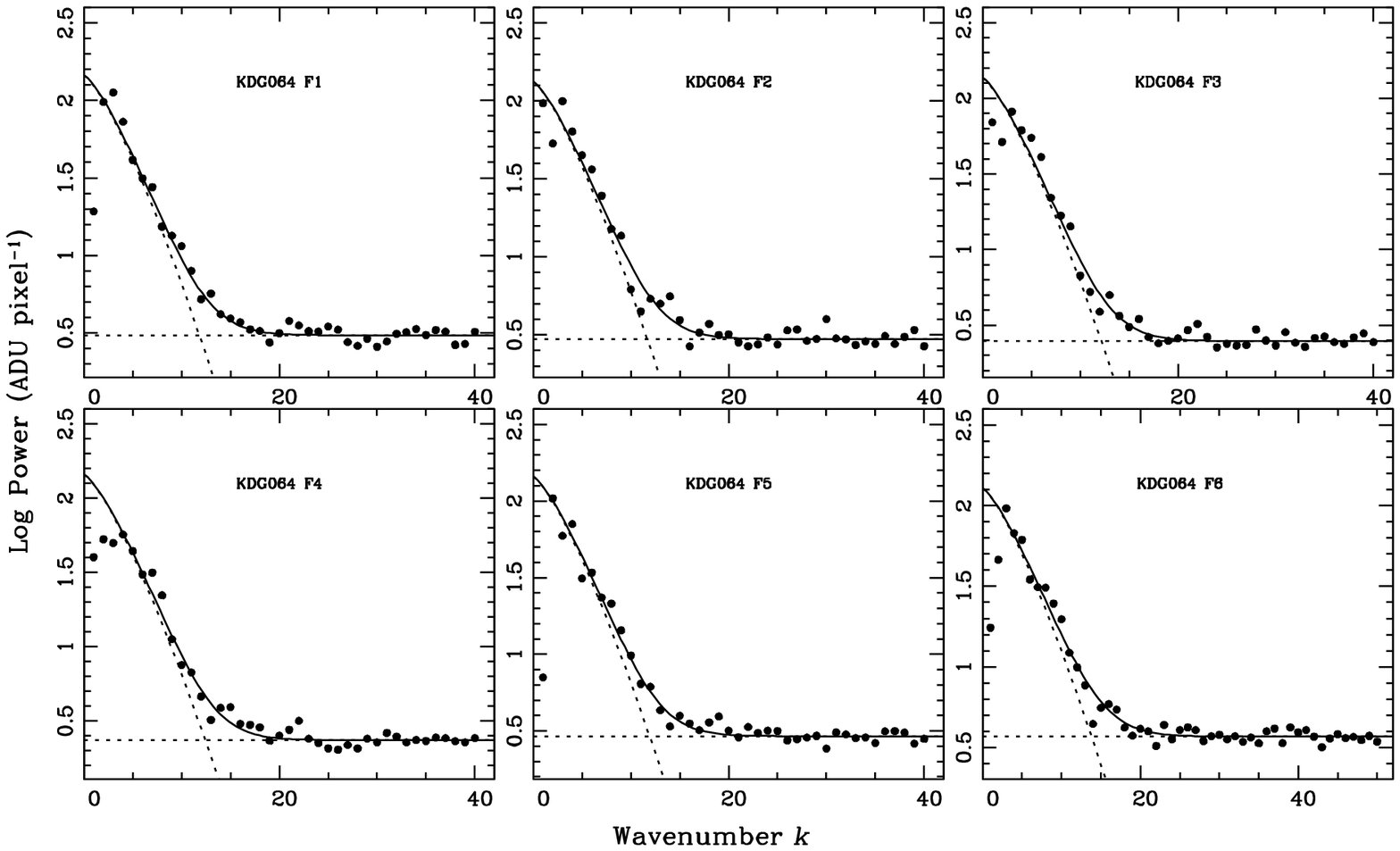}
\includegraphics[height=6.5cm]{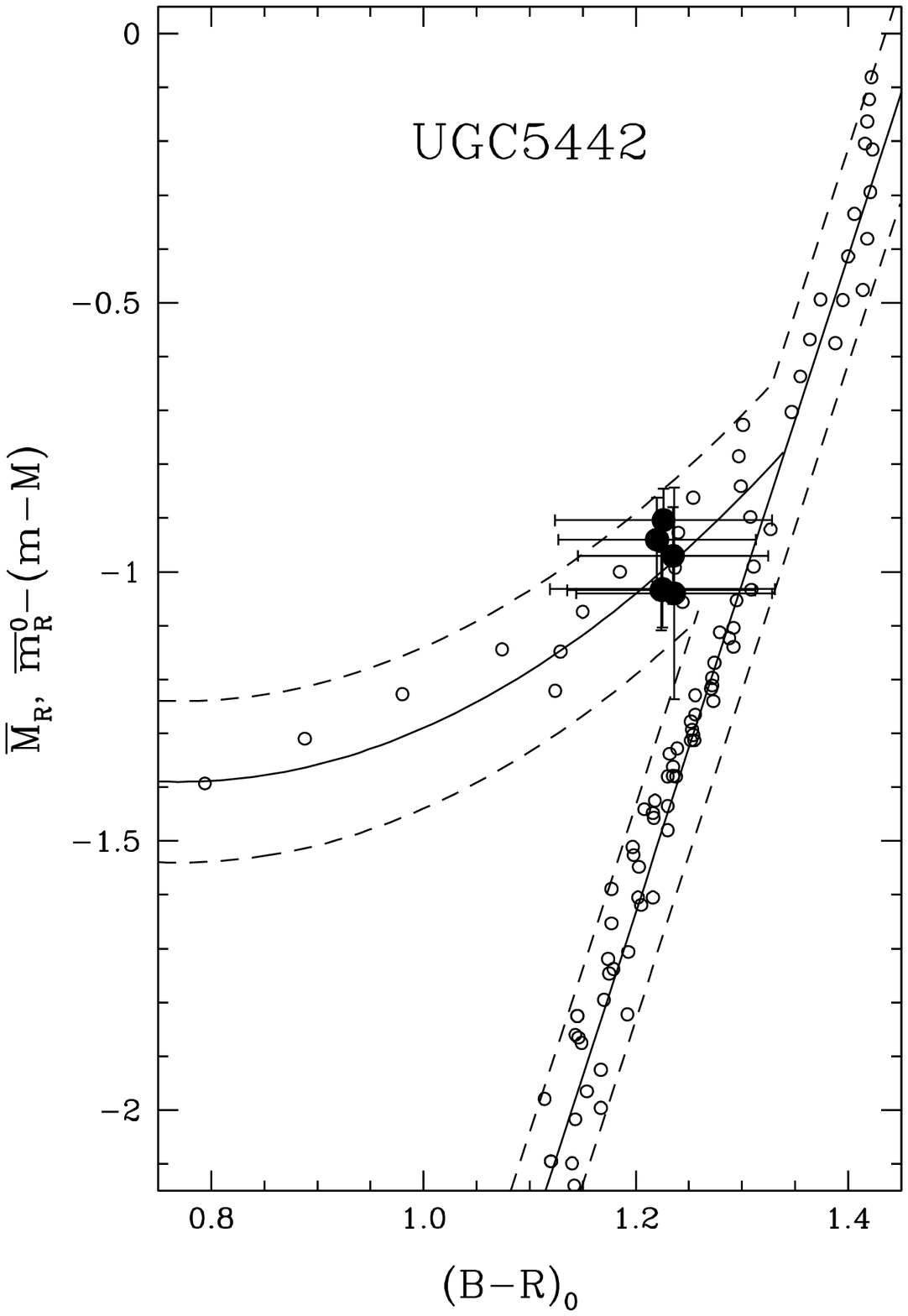}
\caption{
Left: Six fields were selected for the SBF analysis in UGC 5442. The power 
spectra of the SBF fields (filled circles) are well fitted by the sum (solid 
line) of a scaled version of the power spectrum of the PSF and a constant 
(dashed lines). The wavenumbers 1--4 were not considered for the fit. 
Right: Six fields were analysed to derive the distance of UGC 5442. A shift 
by 27.74\,mag yields the best fit of the data to the calibration diagram. 
}
\label{ugc5442}
\end{figure*}

UGC 5442 (KDG 64) is a member of the M81 group. Karachentsev et al.~(2000) 
reported a distance modulus of $(m-M) = 27.84 \pm 0.15$\,mag from the 
measurement of the magnitude of the red giant branch tip. The galaxy has 
a heliocentric velocity of $v_{\odot} = -18 \pm 14$\,km\,s$^{-1}$ 
(Simien \& Prugniel~2002). We analysed six independent SBF fields across the 
galaxy's surface (Fig.~\ref{ugc5442}, left panels). The derived mean distance 
modulus of $(m-M)_{\rm{SBF}} = 27.74 \pm 0.20$\,mag (see Fig.~\ref{ugc5442}, 
right panel) is in good agreement with the TRGB result. Due to the small 
colour range covered by the SBF fields (see Table \ref{flucmagcol}), another 
SBF distance is technically possible by moving the data points onto the linear 
branch of the calibration curve. However, the alternative distance modulus of 
$(m-M)_{\rm{SBF}} = 28.21 \pm 0.3$\,mag is inconsistent with the TRGB result. 

\subsection*{UGC 5944 ([FS90] 047)}
Neither a velocity nor a distance were known for UGC 5944 to date. This dwarf 
elliptical galaxy was catalogued as member of the Leo Group in Ferguson \& 
Sandage (1990) based on its morphology. With our SBF distance of 
$11.1 \pm 0.9$\,Mpc we can now confirm this impression. The derived SBF 
distance is also in good agreement with the mean distance of 11\,Mpc for the 
group (Trentham \& Tully 2002). We note that there was not much of colour 
range found in the five fields we analysed to unambiguously apply the 
distance calibration. Consequently, there is an alternative distance of 
$\approx$22\,Mpc for this galaxy. However, this latter distance appears 
less likely.

\subsection*{NGC 4150}
\begin{figure*}
\centering
\includegraphics[height=6.5cm]{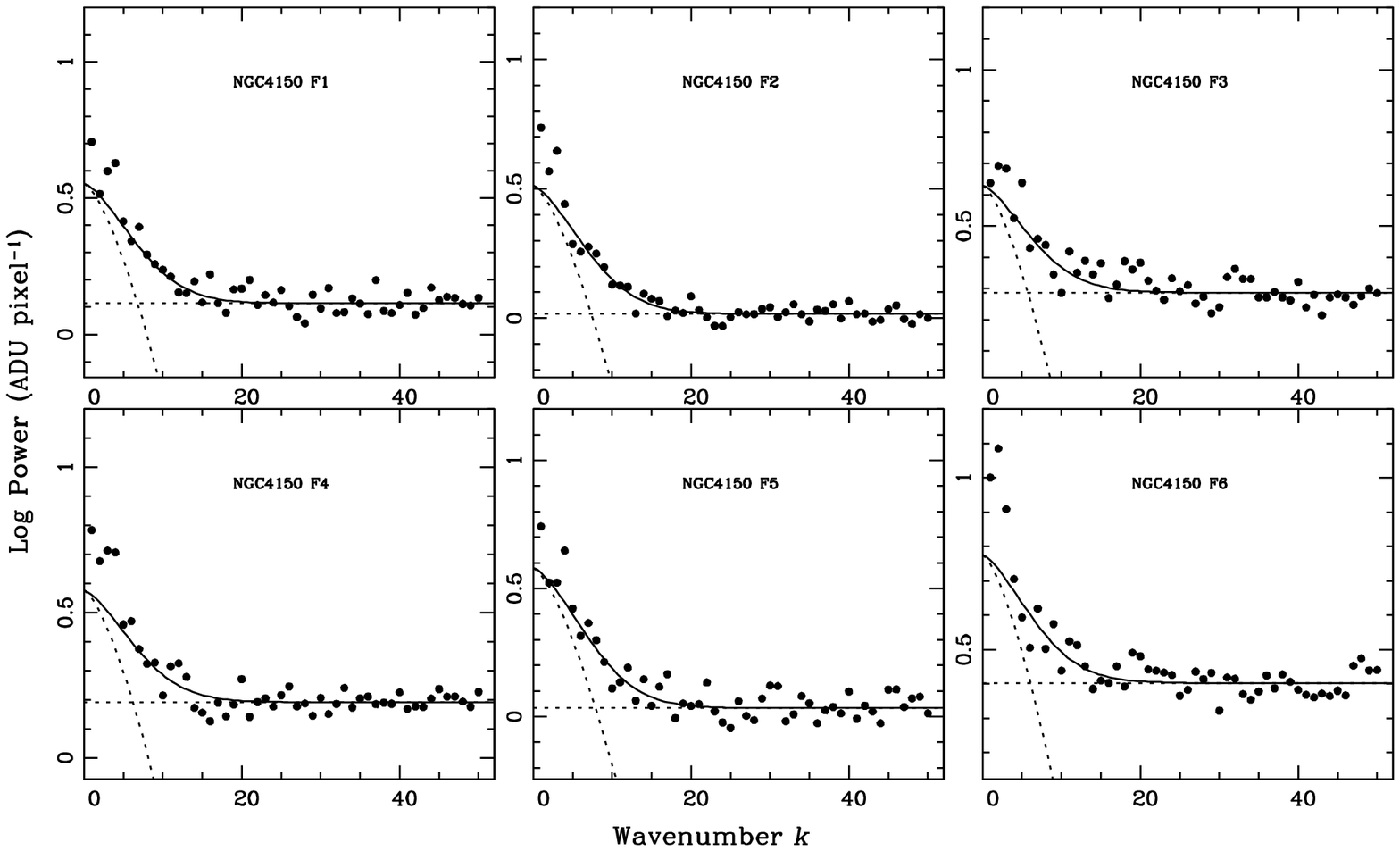}
\includegraphics[height=6.5cm]{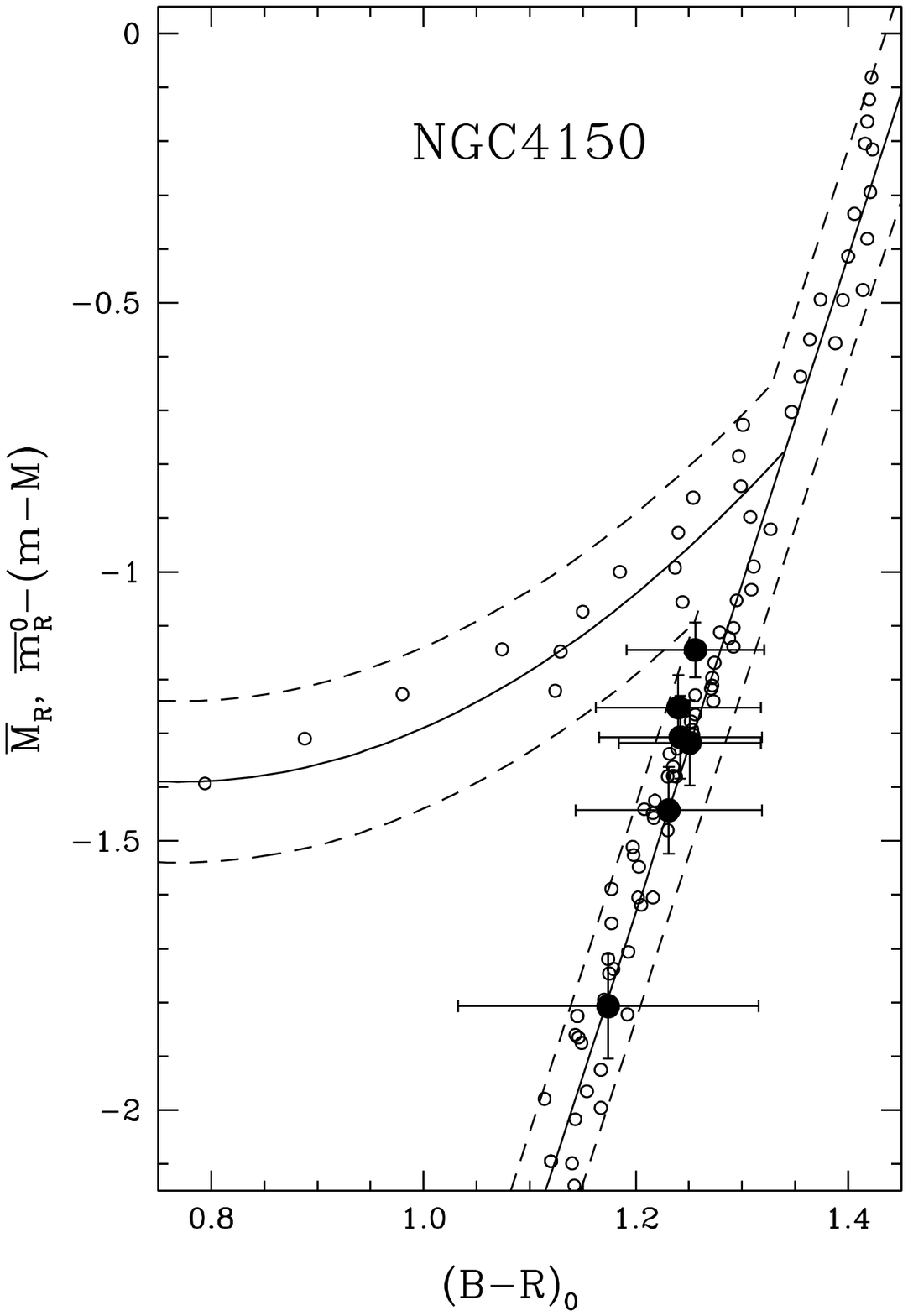}
\caption{
Left: Six fields were selected for the SBF analysis in NGC 4150. The 
signal-to-noise in the power spectra is not as high as measured in the other 
galaxies, as it is more distant. Nevertheless, the power spectra of the SBF 
fields (filled circles) are well fitted by the sum (solid line) of a scaled 
version of the power spectrum of the PSF and a constant (dashed lines). The 
wavenumbers 1--4 were not considered for the fit. 
Right: Six fields were analysed to derive the distance of NGC 4150. A shift 
by 30.79\,mag yields the best fit of the data to the calibration diagram. 
\label{ngc4150res}
}
\label{ngc4150}
\end{figure*}

NGC 4150 is morphologically classified as S0/Sa galaxy (Sandage \& Bedke 
1994). Once the galaxy light model is subtracted, the residual image shows a 
prominent, well developed 4-armed spiral structure with non-circular dust 
pattern in the central region (Fig.~\ref{ngc4150residual}). The galaxy has a 
heliocentric velocity of $v_{\odot} = 226 \pm 22$\,km\,s$^{-1}$ (Fisher et 
al.~1995). Tonry et al. (2001) reported a first SBF distance modulus of 
$(m-M) = 30.69 \pm 0.25$\,mag (or $13.7 \pm 1.7 Mpc$) and Jensen et al. (2003) 
another, corrected SBF distance modulus of $(m-M) = 30.53 \pm 0.24$\,mag (or 
$12.8 \pm 1.5$\,Mpc). We analysed six independent SBF fields across the 
galaxy's surface but well away from the spiral structures 
(Fig.~\ref{ngc4150residual}). The derived distance modulus of 
$(m-M)_{\rm{SBF}} = 30.79 \pm 0.20$\,mag (see Fig.~\ref{ngc4150}) agrees 
with the earlier published SBF results. 

\subsection*{BTS 128}
This dwarf elliptical is located right in the densest region of the Coma I 
group (Binggeli et al.~1990; Trentham \& Tully 2002) about 10 degrees away 
from the northern boundary of the Virgo cluster. The heliocentric velocity of 
BTS 128 of $v_\odot = 1139 \pm 86$\,km\,s$^{-1}$ (Wegner et al.~2001) is in 
good agreement with the Coma I cluster. BTS 128 is 27.5 arcmin away from the 
E0 galaxy NGC 4283 for which a SBF distance of $(m-M) = 30.98 \pm 0.19$\,mag 
was reported by Tonry et al.~(2001). The SBF distance module we derive for BTS 
128 is $(m-M)_{\rm{SBF}} = 31.02 \pm 0.25$\,mag ($16.0 \pm 2.0$\,Mpc). 

\subsection*{UGC 7639}
\begin{figure}[h]
\centering
\includegraphics[width=6.0cm]{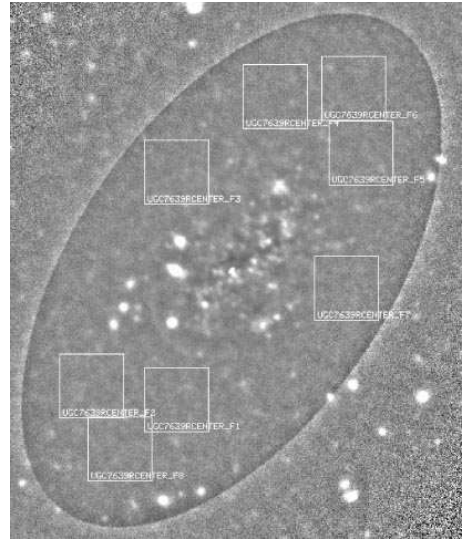}
\caption{
Residual R-band image of UGC 7639 with the eight SBF fields indicated. }
\label{ugc7639cen}
\end{figure}

\begin{figure*}[h]
\centering
\includegraphics[height=6.5cm]{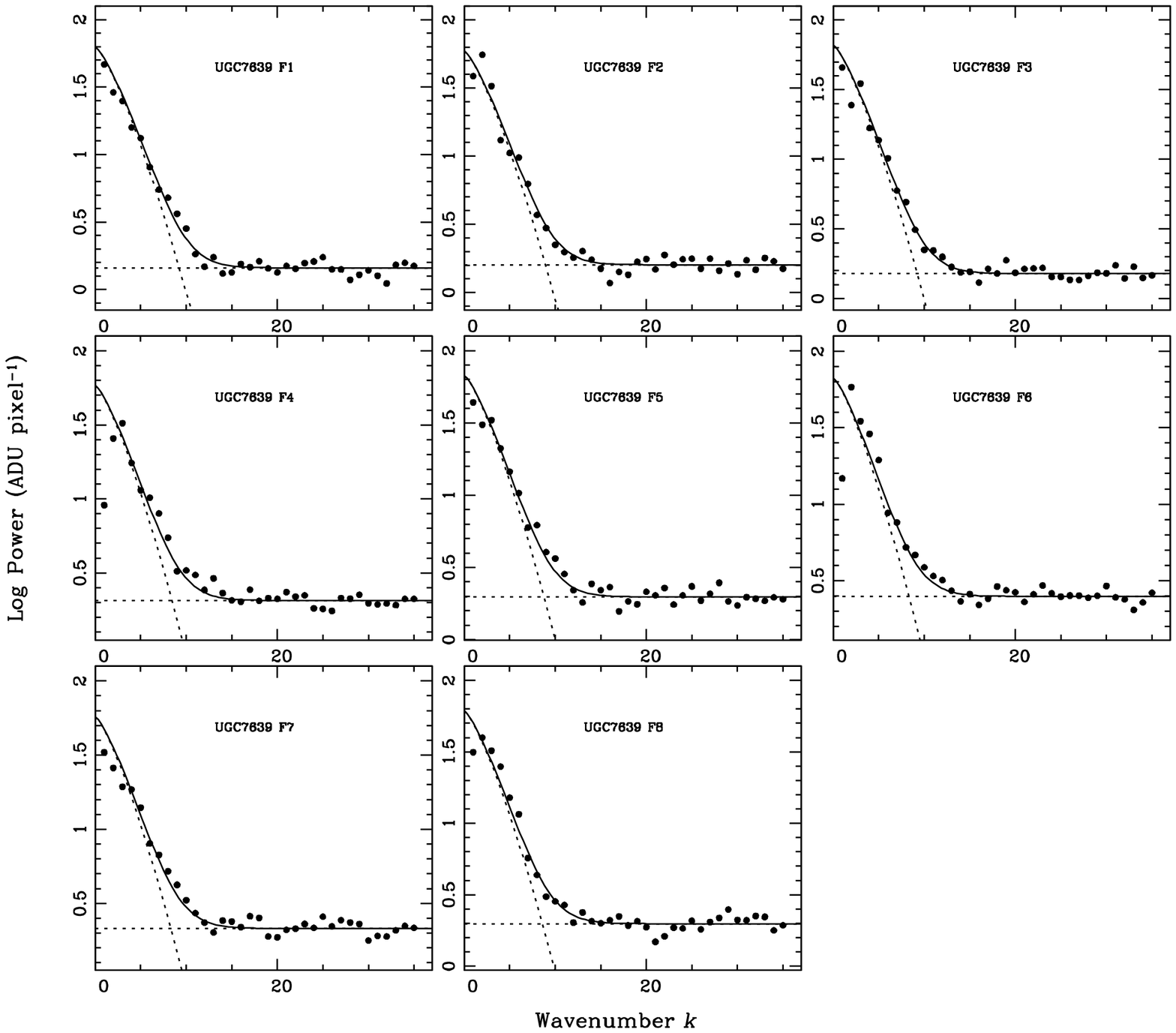}
\includegraphics[height=6.5cm]{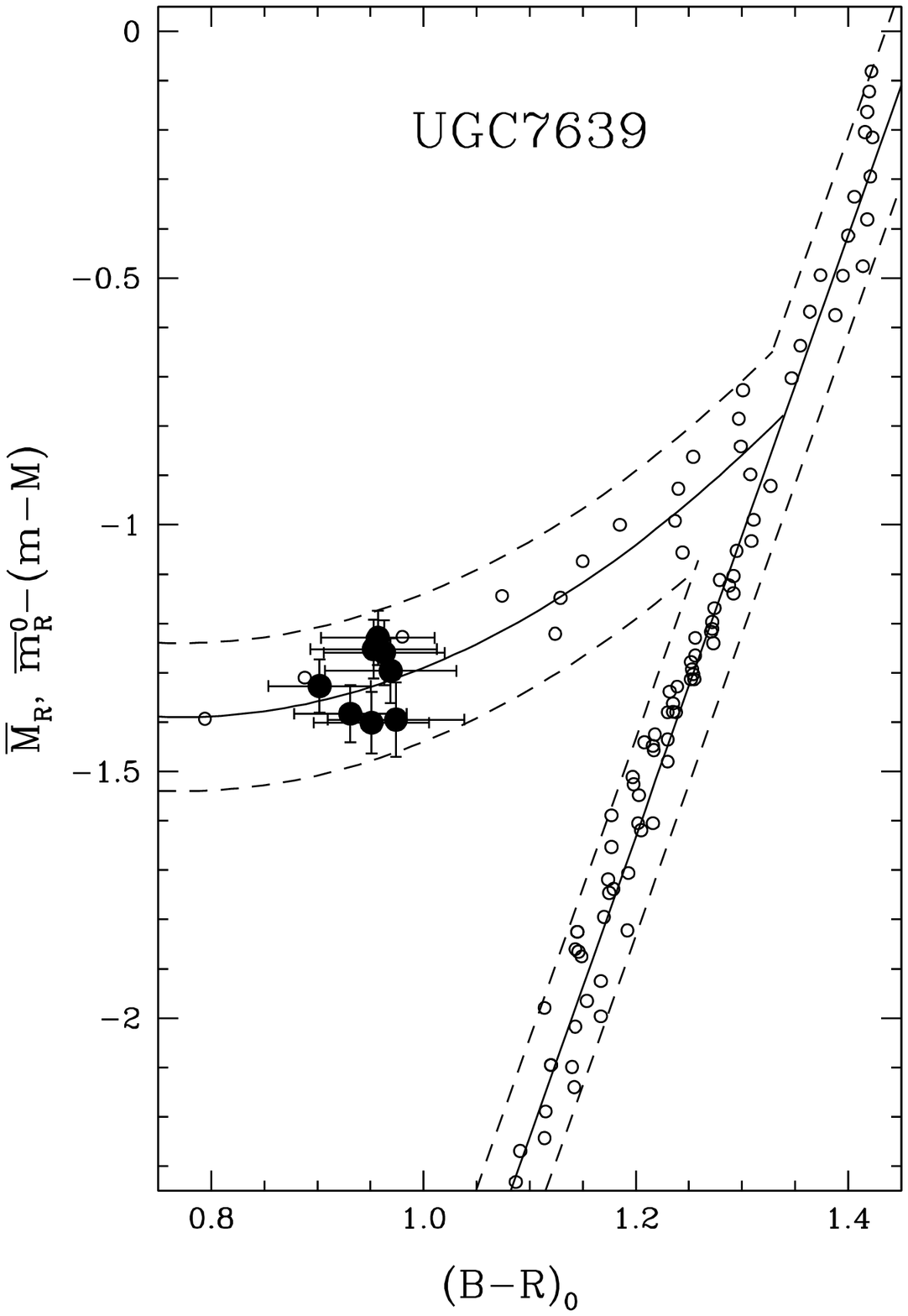}
\caption{
Left: Eight fields were selected for the SBF analysis in UGC 7639. The 
signal-to-noise in the power spectra is high. The power spectra of the SBF 
fields (filled circles) are well fitted by the sum (solid line) of a scaled 
version of the power spectrum of the PSF and a constant (dashed lines). The 
wavenumbers 1--4 were not considered for the fit. 
Right: Eight fields were analysed to derive the distance of UGC 7639. A shift 
by 29.27\,mag yields the best fit of the data to the calibration diagram. 
}
\label{ugc7639}
\end{figure*}

This galaxy, classified as dE/Im, is located in the Canes Venatici region. The 
Im nature of the otherwise featureless dwarf galaxy becomes visible once the 
galaxy model is subtracted, with a central region of bright young stars and 
evidence of dust (see Fig.~\ref{ugc7639cen}). It has a heliocentric velocity 
of $v_\odot = 382$\,km\,s$^{-1}$ (de Vaucouleurs et al. 1991). Makarova et 
al.~(1998) published a rough estimate of the distance of 8.0\,Mpc based on 
brightest blue ($B-V < 0.4$) and red ($B-V > 1.6$) stars. We analysed eight 
independent SBF fields across the galaxy's surface (Fig.~\ref{ugc7639cen} 
and \ref{ugc7639}, left panels). Because of the absence of a steep correlation 
between colour and SBF magnitude for the different SBF  fields, we can use 
unambiguously the parabolic branch of the calibration diagram. The resulting 
SBF distance of $7.1 \pm 0.6$\,Mpc or $(m-M)_{\rm{SBF}} = 29.27 \pm 0.16$\,mag 
(see Fig.~\ref{ugc7639}, right panel) identifies this galaxy as slightly 
closer than previously thought.

\subsection*{UGC 8799}
\begin{figure*}[h]
\centering
\includegraphics[height=5.5cm]{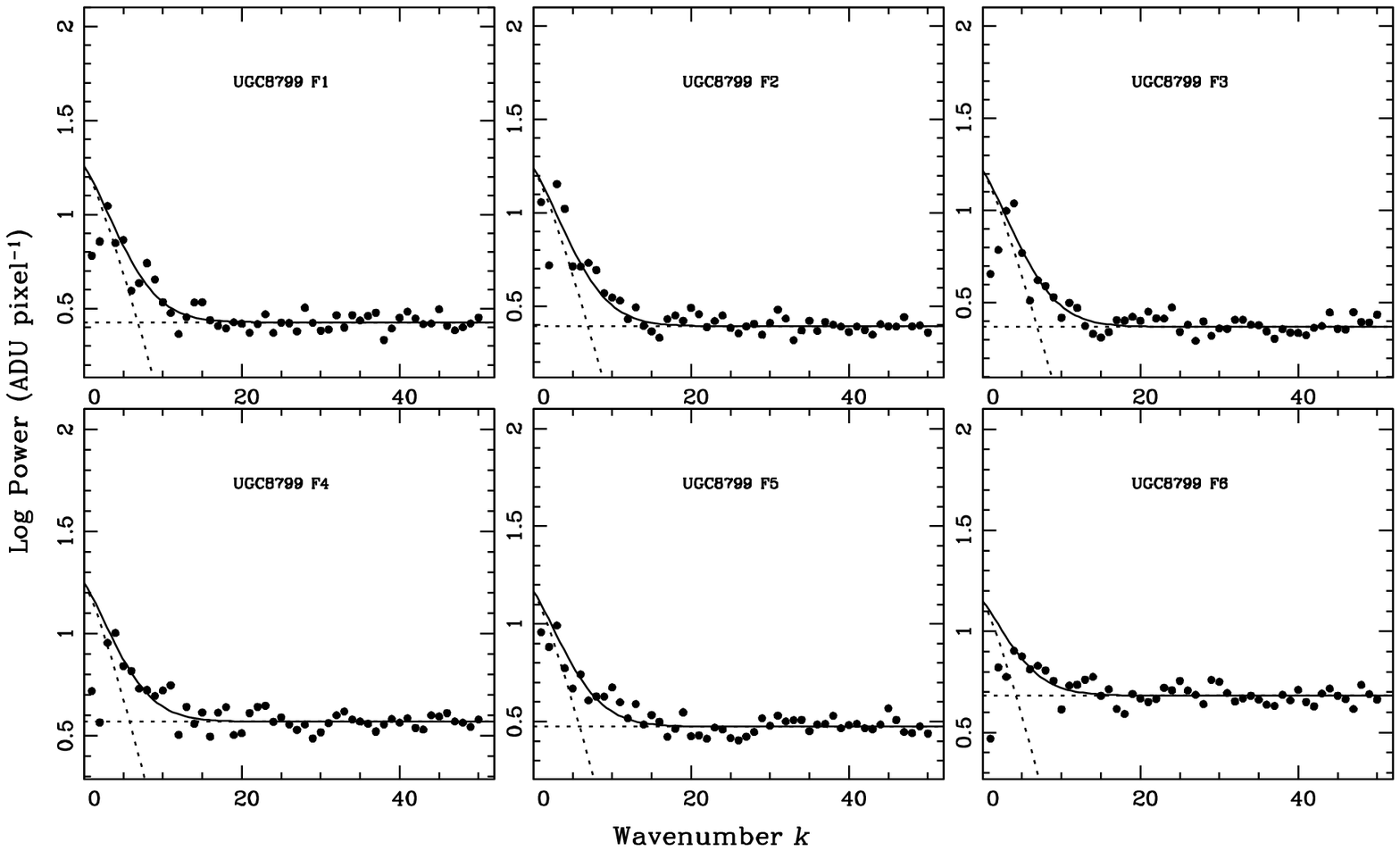}
\includegraphics[height=5.5cm]{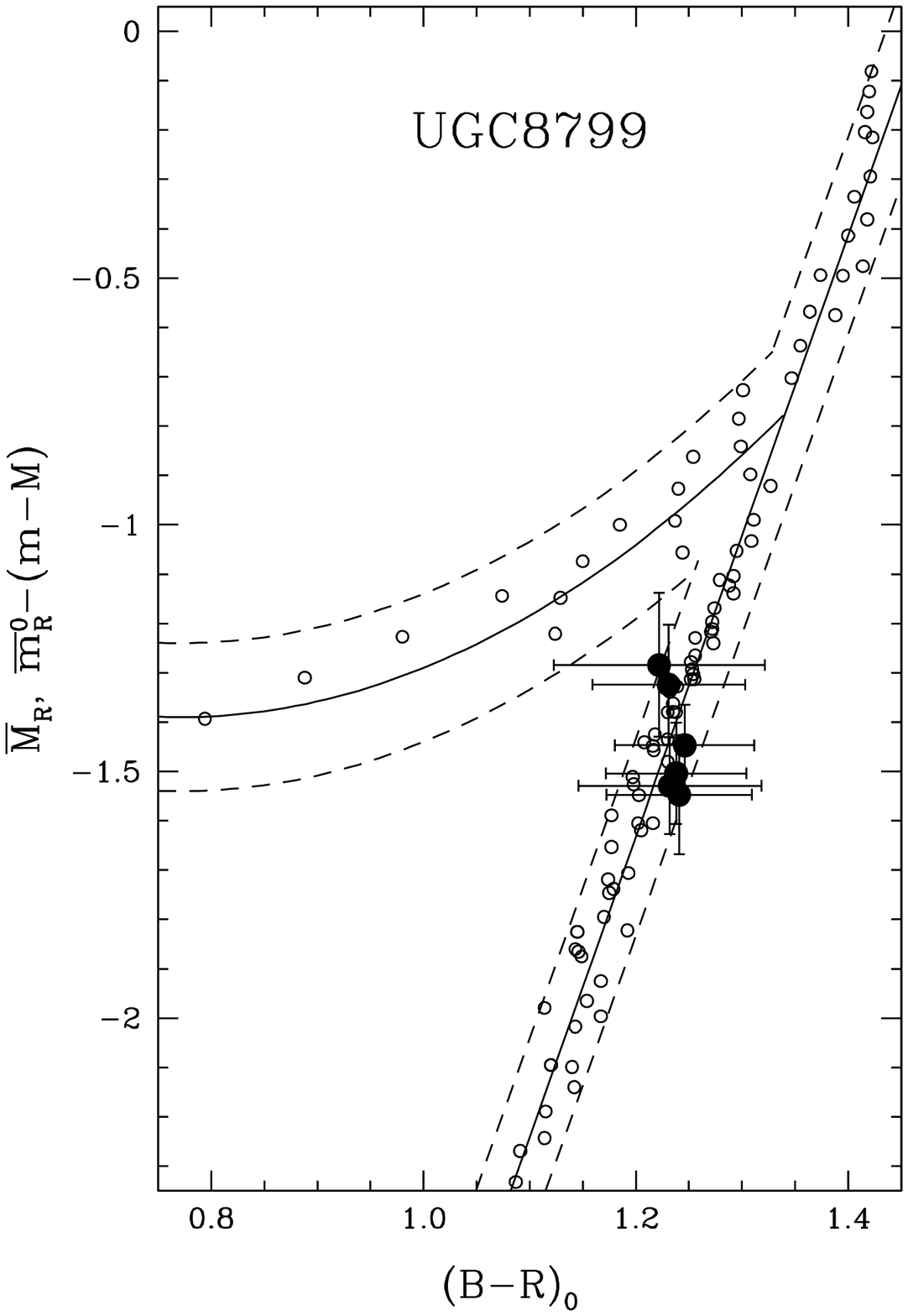}
\includegraphics[height=5.5cm]{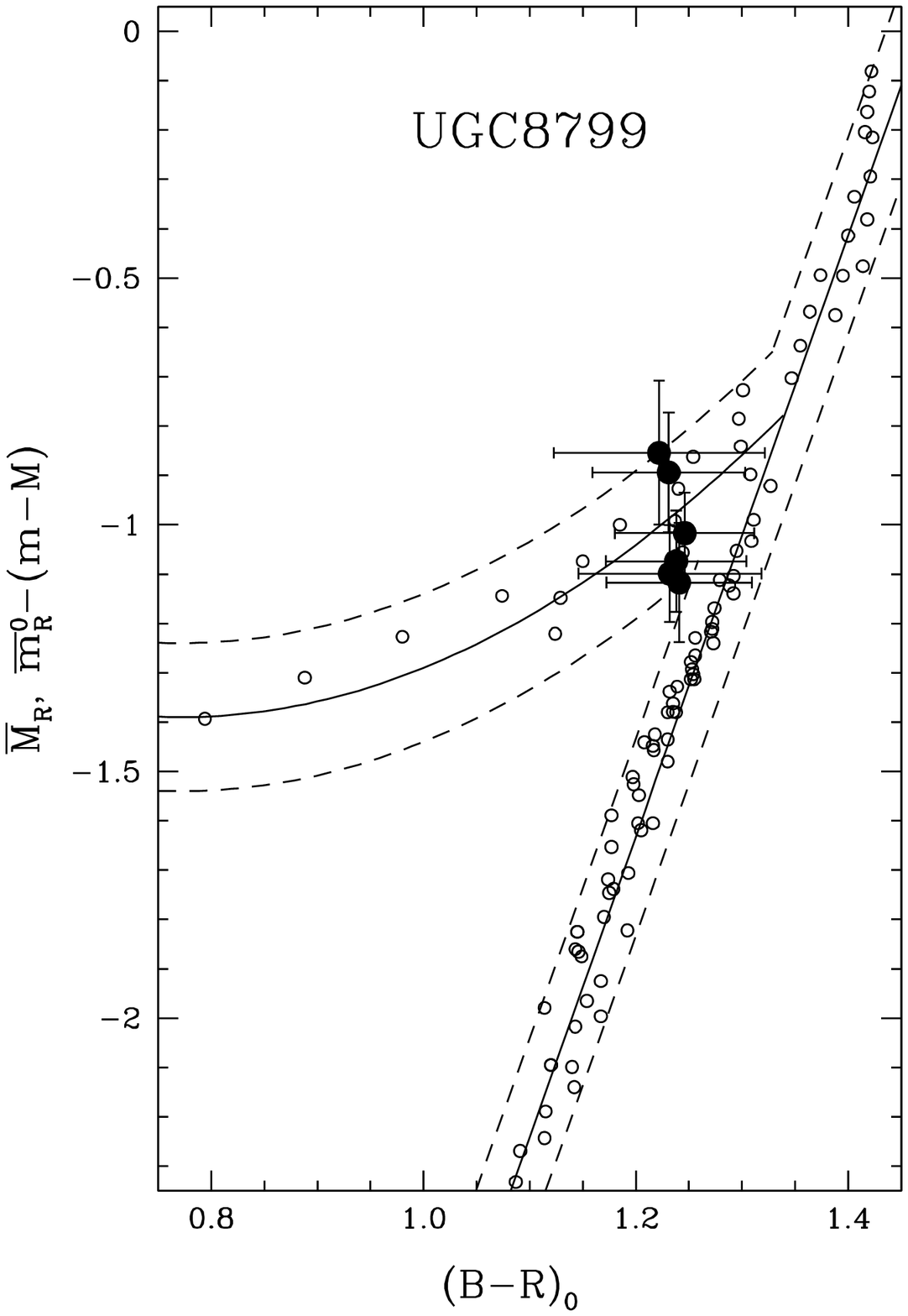}
\caption{Left: Six fields were selected for the SBF analysis in UGC 8799. The 
signal-to-noise in the power spectra is not as high as measured in the other 
galaxies due to the larger distance of the galaxy at $\sim 13.2$\,Mpc. 
Nevertheless, the power spectra of the SBF fields (filled circles) are 
reasonably well fitted by the sum (solid line) of a scaled version of the 
power spectrum of the PSF and a constant (dashed lines). The wavenumbers 
1--4 were not considered for the fit.
Right: Six fields were analysed to derive the distance of UGC 8799. A shift 
by 30.19\,mag or 30.61\,mag yields the best fit of the data to the calibration 
diagram. 
}
\label{ugc8799}
\end{figure*}

No distance was known for this dwarf elliptical to date but Geller et 
al.~(1997) and Wegner et al.~(2001) report a velocity of 1132\,km\,s$^{-1}$. 
Because this galaxy is only 1.5 hours in R.A. from the western boundary of the 
Virgo cluster and the velocity agrees well with the mean cluster velocity 
($980 \pm 60$\,km\,s$^{-1}$, Tanaka 1985) this may suggest that UGC 8799 has a 
distance that falls into the observed distance range of the Virgo cluster 
14\,Mpc $<$ D $<$ 23\,Mpc (Jerjen et al.~2004). The six SBF fields (see 
Fig.~\ref{ugc8799}) show only a narrow $B-R$ colour spread and the 
signal-to-noise in the power spectra is low. This leads to two possible 
distance moduli:  $(m-M)_{\rm{SBF}} = 30.19 \pm 0.23$\,mag or 
$(m-M)_{\rm{SBF}} = 30.61 \pm 0.26$\,mag (see Fig.~\ref{ugc8799}), suggesting 
it is in the outskirts of the Virgo cluster. 

\subsection*{UGC 8882}
\begin{figure*}[h]
\centering
\includegraphics[height=6.5cm]{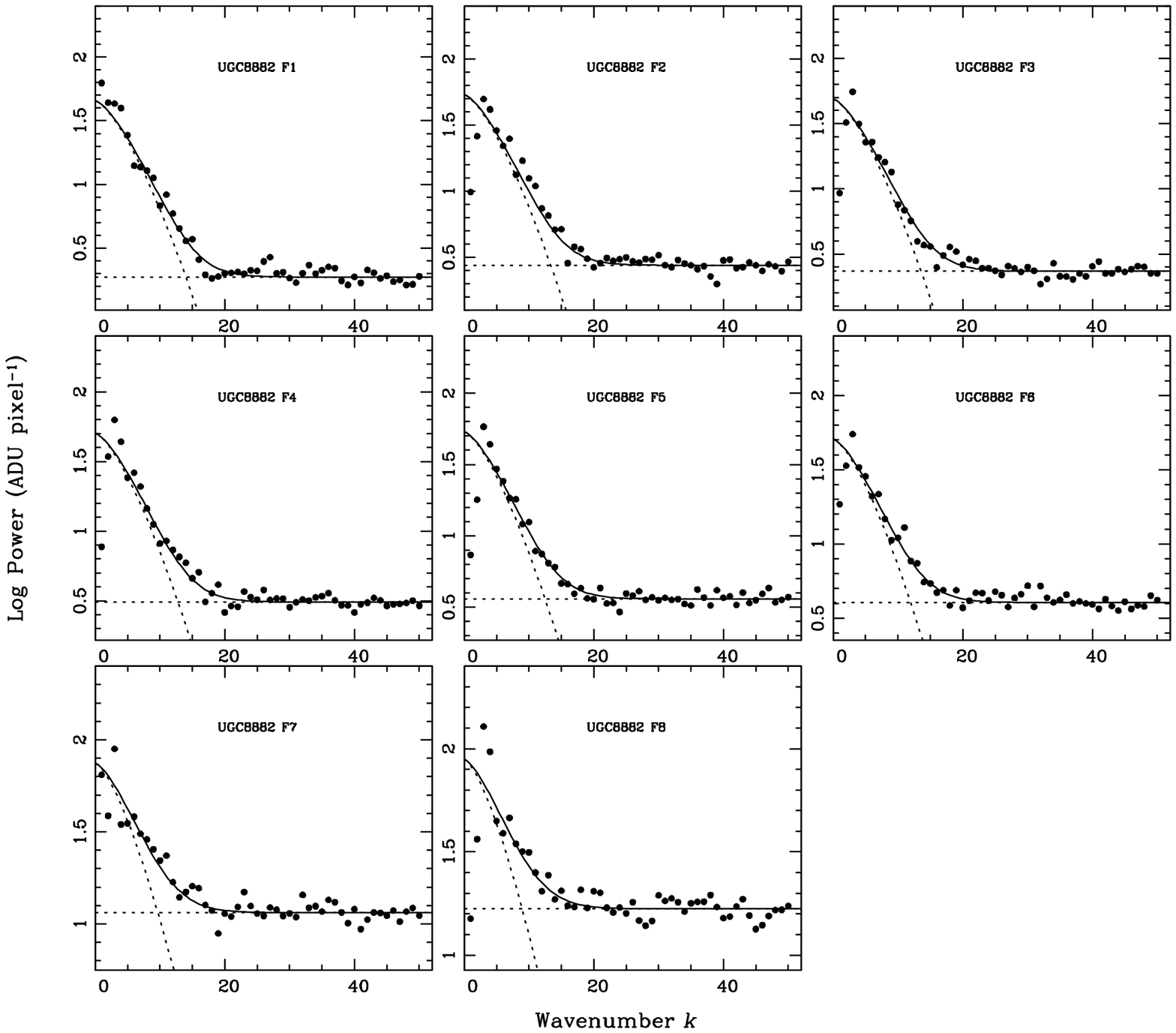}
\includegraphics[height=6.5cm]{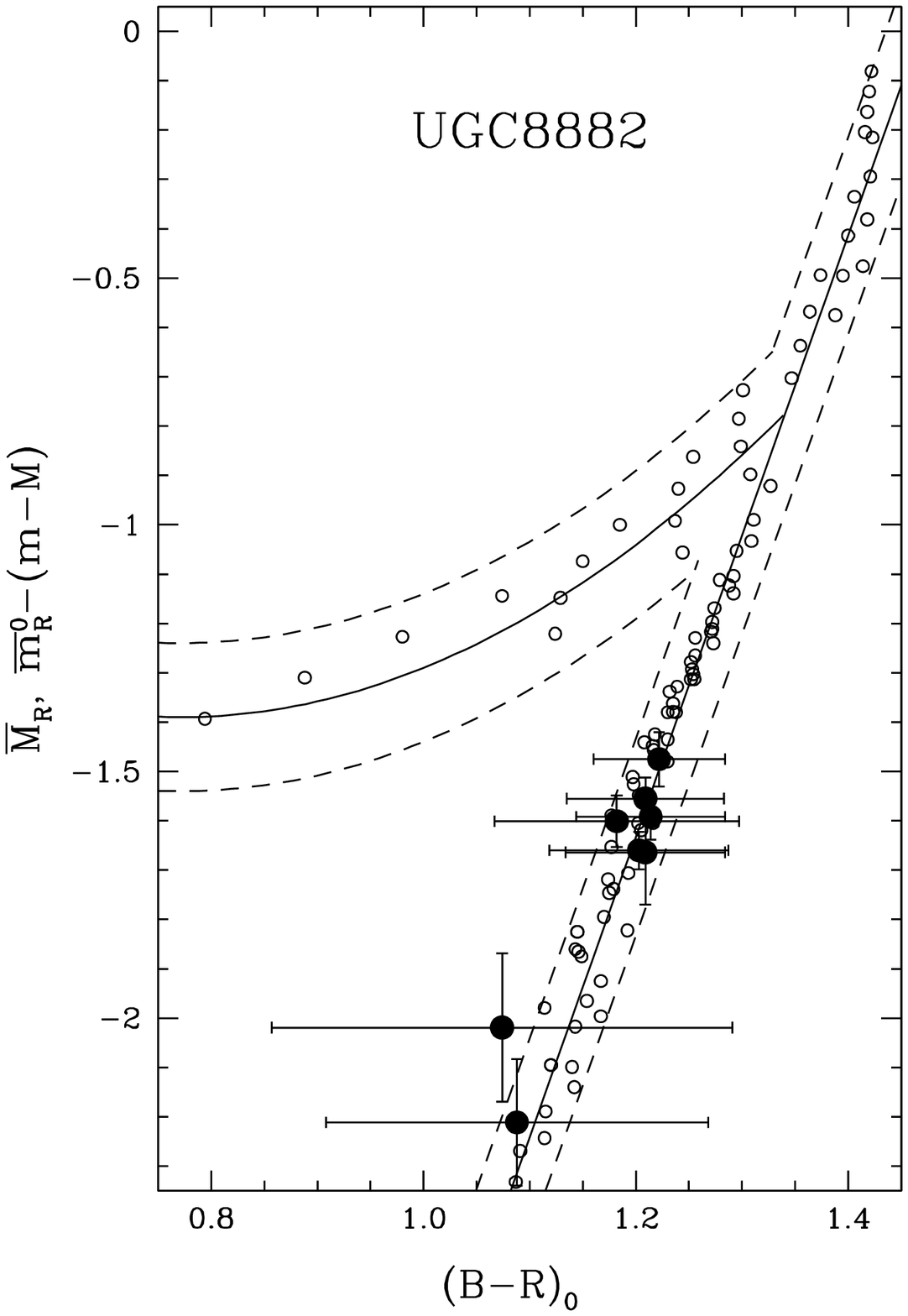}
\caption{
Left: Eight fields were selected for the SBF analysis in UGC 8882. The 
signal-to-noise in the power spectra is not as high as measured in the other 
galaxies due to the larger distance of the galaxy at 14.4\,Mpc. Nevertheless, 
the power spectra of the SBF fields (filled circles) are well fitted by the 
sum (solid line) of a scaled version of the power spectrum of the PSF and a 
constant (dashed lines). The wavenumbers 1--4 were not considered for the fit. 
Right: Eight fields were analysed to derive the distance of UGC 8882. A shift 
by 29.60\,mag yields the best fit of the data to the calibration diagram. 
}
\label{ugc8882}
\end{figure*}

Neither a velocity nor a distance was available for this nucleated dwarf 
elliptical galaxy. Bremnes et al.~(1999) claim the galaxy is part of the M101 
group and reported a mean $B-R$ colour of 1.29. The Cepheid based distance 
modulus for M101 is $(m-M)_0 = 29.28 \pm 0.14$\,mag (Stetson et al.~1998 and 
references therein). We analysed eight independent SBF fields across 
the galaxy's surface (Fig.~\ref{ugc8882}, left panels). Because of the 
observed linear relation between colour and SBF magnitude for the different 
fields we used the linear branch of the calibration diagram. The derived 
distance modulus is $(m-M)_{\rm{SBF}} = 29.60 \pm 0.20$\,mag 
(see Fig.~\ref{ugc8882}, right panel). This result is consistent with UGC 8882 
being a member of the M101 group.

\section{Summary and Conclusions}
We have analysed $BR$-band CCD images of nine nearby dwarf ellipticals 
and one S0 galaxy. We have measured their stellar $R$-band 
surface brightness fluctuation magnitudes $\overline{m}_R$ and $(B-R)_0$ 
colours in 61 galaxy fields. The resulting distances were compared with 
existing distances measured with the TRGB method and SBF method. 
Agreement between our distances and those of TRGB and previous SBF 
measurements was very high; all distances agreed within 0.1 magnitudes. 

Our SBF distances are given in Table \ref{distmods} together with 
heliocentric radial velocities and radial velocities relative to the 
centre of mass of the Local Group of galaxies. The latter were calculated 
from heliocentric radial velocities following de Vaucouleurs et al.~(1991) 
and using our distances to the galaxies. The Local Group centre of 
mass was assumed to lie exactly half way between the Milky Way 
and M31. The SBF distance to velocity relation positions BTS 128, UGC 7639 
and UGC 8799 quite well on the Hubble flow. This is not the case with 
NGC 4150, KDG 61 and UGC 5442. 

Our SBF distance to NGC 4150 of $14.4 \pm 0.7$ Mpc is in good agreement
with earlier SBF distances (Tonry et al.~2001, Jensen et al.~2003) of 
$13.7\pm1.7$\,Mpc and $12.8\pm1.5$\,Mpc, and places this galaxy almost 10 Mpc 
behind the centre of Canes Venatici I cloud. All these estimates are 
consistent with Karachentsev et al.~(2003) lower distance limit of 6.3 Mpc 
using the TRGB method in a study of CVn I cloud, and their very rough 
estimate of 20 Mpc, using the globular cluster luminosity function method. 
The galaxy might then be associated with the outskirts of the Virgo cluster. 
The velocity of NGC 4150 is rather low ($v_{\odot} = 226 $km\,s$^{-1}$), but 
it is difficult to say whether it is discordant with either our distance
estimate of $14.4 \pm 0.7$ Mpc or Karachentsev et al.'s~(2003) $\approx
20 $Mpc. Solanes et al.~(2002) have found that most Virgo galaxies in
the region closest to us, corresponding to our distance to NGC 4150,
have high radial velocities outward from the cluster centre (with
similarly unusually high velocities away from us for Virgo cluster
galaxies in the region behind the centre of the cluster). The angular
separation of NGC 4150 from the cluster centre is 19 degrees. If it has
a high outward velocity from the cluster centre, its line-of-sight
velocity could be quite low. The observed velocity of NGC 4150 can be 
consistent with membership in the Virgo cluster, at its outer edge. 

Our SBF distance to KDG 61 confirms the existing TRGB distance (Karachentsev 
et al.~2000) and thus the membership in M81 group. Another M81 group member is 
confirmed as our SBF distance to UGC 5442 agrees well with a previous TRGB 
distance (Karachentsev et al.~2000). Radial velocities of KDG 61 and UGC 5442 
are reasonable relative to M81 velocity of $v_{\odot} = -34 \pm 4$ (de 
Vaucouleurs et al.~1991) or $v_{LG} \approx 96$ km\,s$^{-1}$. 

UGC 7639 had only a tentative distance measured with brightest blue and red 
stars (Makarova et al.~1998) before our SBF distance, which is in relatively 
good agreement with the earlier distance and confirms the location of this 
galaxy in the Canes Venatici II cloud. 

Distances to galaxies UGC 1703, UGCA 200, UGC 5944 and UGC 8882 were not 
previously known. We have provided accurate SBF distances to these galaxies. 
Our distances suggest UGC 1703 may be a distant companion of NGC 784, 
UGCA 200 may not be a companion of NGC 3115 as was assumed before, 
UGC 5944 is most certainly a member of Leo I group and UGC 8882 seems to be 
a member of M101 group. Likewise only radial velocities were known for 
galaxies BTS 128 and UGC 8799 before our SBF distances, which confirm the 
membership of BTS 128 in Coma I group and suggest UGC 8799 lies at the 
outskirts of Virgo I cluster. Its angular separation from the cluster centre, 
at M87 location of 12$^h$ 31$^m$ in RA and +12$^\circ$ 23$^m$ in 
dec. (J2000.0), is 21 degrees. 

The SBF distances we have presented continue to support the understanding of 
the distribution of dwarf galaxies in galaxy groups and intermediate space in 
the Local Group neighbourhood. They also demonstrate well the feasibility of 
the Surface Brightness Fluctuation method in determining accurate distances 
with 2m class ground-based telescopes out to the near side of the Virgo 
cluster. 

\begin{table}
\caption{SBF distances of the sample galaxies with the ambiguous case listed 
twice. Heliocentric and Local Group barycentric velocities are also listed if 
available. 
\label{distmods}}
\begin{tabular}{lcccc}
\hline\hline
Galaxy   &  $(m-M)_0$      &  $D$           &  $v_{\odot}$  & $v_{LG}$\\
Name     &  (mag)          &  (Mpc)         &  (km\,s$^{-1}$)       & (km\,s$^{-1}$) \\
 (1)     &  (2)            &  (3)           &  (4)          & (5) \\
\hline
UGC 1703 &  28.11$\pm$0.15 &  4.2 $\pm$ 0.3 &    -- &   -- \\
KDG 61   &  27.61$\pm$0.17 &  3.3 $\pm$ 0.3 &  -135 &   -4 \\ 
UGCA 200 &  29.01$\pm$0.27 &  6.3 $\pm$ 0.8 &    -- &   -- \\
UGC 5442 &  27.74$\pm$0.18 &  3.5 $\pm$ 0.3 &   -18 &  106 \\
UGC 5944 &  30.22$\pm$0.17 & 11.1 $\pm$ 0.9 &    -- &   -- \\
NGC 4150 &  30.79$\pm$0.11 & 14.4 $\pm$ 0.7 &   226 &  207 \\
BTS 128  &  31.02$\pm$0.25 & 16.0 $\pm$ 1.9 &  1139 & 1129 \\
UGC 7639 &  29.27$\pm$0.16 &  7.1 $\pm$ 0.6 &   382 &  447 \\
UGC 8799 &  30.19$\pm$0.23 & 10.9 $\pm$ 1.2 &  1132 & 1094 \\
 ---     &  30.61$\pm$0.26 & 13.2 $\pm$ 1.7 &       &      \\
UGC 8882 &  29.60$\pm$0.20 &  8.3 $\pm$ 0.8 &    -- &   -- \\
\hline
\hline
\end{tabular}
\end{table}

\begin{acknowledgements}
The Nordic Optical Telescope is operated on the island of La Palma 
jointly by Denmark, Finland, Iceland, Norway, and Sweden, in the 
Spanish Observatorio del Roque de los Muchachos of the Instituto de 
Astrof\'isica de Canarias. We thank Kari Nilsson for his help 
with the observations and the Academy of Finland for support 
through its funding of the ANTARES programme. Financial support for RR 
has been provided by Finnish Graduate School in Space Physics and Astronomy 
and by the Academy of Finland through funding of the project 
``Calculation of Orbits''. 
This research has made use of the NASA/IPAC Extragalactic Database (NED), 
which is operated by the Jet Propulsion Laboratory, California Institute of 
Technology, under contract with the National Aeronautics and Space 
Administration. 
We are grateful to the referee Enzo Brocato for his useful comments.
\end{acknowledgements}

\end{document}